 \documentclass[10pt,journal,final,letterpaper,twocolumn,twoside]{IEEEtran}
\usepackage{amsmath,amsfonts,bm,amssymb,psfrag,ifthen,color,subfigure}
\usepackage{mathrsfs}
\usepackage[justification=centering,font=footnotesize]{caption}
\usepackage{todonotes}

\allowdisplaybreaks[4]

\usepackage[dvips]{epsfig}
\usepackage[dvips]{graphicx}
\usepackage{amsmath,amsfonts,bm,amssymb,psfrag,ifthen,color,subfigure}
\usepackage{algpseudocode}
\usepackage{algorithm}
\usepackage{algorithmicx}
\usepackage{ifthen}
\usepackage{setspace}
\usepackage{cite}
\usepackage{url}
\usepackage{longtable}
\usepackage{stfloats}  % Written by Sigitas Tolusis
\usepackage{graphics,booktabs,color,subfigure}
\usepackage{float}

\newcommand{\tabincell}[2]{\begin{tabular}{@{}#1@{}}#2\end{tabular}}

\usepackage{listings}

\floatname{algorithm}{Algorithm}

\begin{document}
\title{Optimal Power Allocation for  Integrated  Visible Light Positioning and Communication   System with a Single LED-Lamp}

\author{Shuai Ma, Ruixin Yang, Bing Li, Yongyan Chen, Hang Li,  Youlong Wu, Majid Safari, Shiyin Li, and Naofal Al-Dhahir

\thanks{S. Ma, R. Yang, B. Li, Y. Chen  and S. Li are with the School of Information and Control Engineering, China
    University of Mining and Technology, Xuzhou, 221116,
    China. (e-mail: mashuai001@cumt.edu.cn).}
}

\maketitle
 \begin{abstract}

 In this paper, we investigate  an integrated visible light positioning and communication (VLPC) system with   a single   LED-lamp.
First, by leveraging the fact that the VLC channel model  is a function of the receiver's location,
we propose a system model that estimates the channel state information (CSI) based on the positioning information without transmitting pilot sequences.
Second, we derive  the Cramer-Rao lower bound (CRLB) on the positioning error variance and a lower bound on the achievable rate with  on-off keying modulation.
Third, based on the derived performance metrics,  we  optimize the   power allocation to minimize the CRLB, while satisfying the rate outage probability constraint.
To tackle this non-convex optimization  problem, we apply  the worst-case distribution of  the Conditional Value-at-Risk (CVaR) and the
 block coordinate descent (BCD) methods
 to obtain the feasible solutions.
Finally, the effects    of critical  system
parameters, such as outage probability,  rate   threshold,  total power threshold, are revealed by numerical results.

\end{abstract}
\begin{IEEEkeywords}
  Visible light communication,   Visible light positioning, Power allocation, Cramer-Rao lower bound.
\end{IEEEkeywords}

\IEEEpeerreviewmaketitle

\section{Introduction}

With the explosively increasing number of Internet of
Things (IoT)  devices in beyond fifth generation (B5G)  networks,
the  crisis of  radio frequency (RF) spectrum  shortage  becomes
increasingly  challenging, which makes it more difficult for  the RF  wireless systems  to meet the high speed   data
transmission  and high accuracy  positioning   demands simultaneously\cite{Tsonev_2015_Towards}.
  It is worth noting that  more than   50$\%$ of voice traffic and
70$\%$ of wireless data  traffic occur in indoor environments \cite{Weichold_2015}.
Since the indoor activity requires both  illumination and  network access,
visible light communication (VLC) \cite{Elgala_2011_Indoor}
and visible
light positioning (VLP)\cite{Keskin_2018_Localization},
which  apply the ubiquitous  light emitting diodes (LEDs)  as access points (APs) and anchor nodes, are  promising  technologies for  indoor IoT applications.
Comparing with the conventional radio frequency (RF)  wireless technologies,  the   distinct advantages of VLC and VLP  are multifold \cite{Keskin_2018_Localization}
 \cite{Pathak_2015_Visible}, including
no electromagnetic interference, high energy efficiency,  high security,  and low-cost.

VLC utilizes  the simple   intensity modulation
and direct detection (IM/DD) mechanism for information
transmission, and has   attracted significant research interests as a breakthrough  technology for B5G  networks \cite{Jovicic_2013_Visible}.
Extensive studies have been reported to  improve VLC networks performance.
For example, by
adopting the alternating direction method of multipliers
(ADMM), a distributed coordinated   interference management scheme is proposed in
\cite{MA_TWC_2019}  for VLC   networks. To balance energy and bandwidth efficiency,
both  power allocation and rate splitting are optimized in \cite{Deng_Access_2019} for DC-biased optical orthogonal frequency  division multiplexing (DCO-OFDM).
In \cite{Gao_TWC_2017}, in order to jointly optimize the post-equalizer, the precoder and the DC  offset,
a gradient projection-based
procedure is presented   to minimize the sum mean squared error (MSE) of  the received symbols.   Furthermore, VLC has been commercialized in industry.
Some startup companies and existing industry giants,
such as  pureLiFi, Philips, and
Oledcomm, are providing VLC commercial solutions  in home and business buildings.

Owning to its short wavelength  and low   multipath interference, VLP can achieve high  indoor positioning accuracy,
which can facilitate various applications,
such as indoor navigation, location aware services, logistic management, and assets tracking, to name few.
By exploiting different visible light characteristics,
existing  VLP schemes can apply
time of arrival (TOA) \cite{2016_Theoretical_Amini}  \cite{Akiyama_Time_2017},
time difference of arrival (TDOA)\cite{Jung_TDOA_2012},
angle of  arrival (AOA)\cite{2014_Three_Yang}
and  received signal strength (RSS)\cite{Yu_IPJ_2018} techniques for positioning.
Among the above VLP schemes, the RSS based scheme is widely adopted
due to its simplicity and ubiquity, where
the distance  between the  lamp  base station (BS) and the device is calculated based on the channel model.
For example,  by using the weighted k-nearest-neighbor (K-NN),  a   multi-LEDs positioning system is designed
in \cite{Bakar_2021_Accurate} based on sparse fingerprints.
In \cite{Du_2019_Experimental}, an artificial neural
network (ANN)-based position
estimator is proposed for 3D RSS-based VLP systems.

Most of the existing works  only focus on VLC or VLP  individually. In practical indoor applications,
an integrated system with both the communication and positioning functions is highly desirable.
So far, only few works considered the integration of VLC and VLP.  Specifically,
VLC systems based on  orthogonal frequency division multiplexing access (OFDMA) \cite{Lin_IPJ_2017,Xu_Accuracy_2017} were proposed to estimate the receiver position.
 An integrated visible
light  positioning and communication (VLPC) system was designed in \cite{Yang_Demonstration_2018}  by using filter
bank multicarrier-based subcarrier multiplexing (FBMC-SCM).
Such FBMC-SCM-assisted VLP with its high signal processing complexity was proposed to reduce the out-of-band interference (OOBI).
Towards an OFDMA VLPC   network,
 the authors in   \cite{Yang_2020_QoS} jointly optimized the AP selection, bandwidth allocation,
adaptive modulation, and power allocation to maximize the
 data rate while satisfying positioning accuracy constraints.
In \cite{Yang_WCL_2019}, a  modified experience replay
actor-critic (MERAC) reinforcement learning (RL) approach was presented to maximize the  sum rate   under the  users' minimum data  rates and positioning accuracy  requirements.
 In \cite{Yang_Coordinated_2020}, the authors proposed a coordinated
resource allocation approach  to maximize the sum rate while satisfying the minimum
data rates and positioning accuracy requirements of devices.

In  the above considered
 VLPC systems, it is  required that  at least
two   lamp signals  are captured at the receiver simultaneously  for effective positioning.
Unfortunately, such multi-lamp setup may not fit in many practical scenarios,  such as in a tunnel,  corridor,  and  staircase, where the lamps are sparsely installed.
 In these scenarios, the multi-lamp based  method  will not be as efficient as in a large and flat room.
In  terms of system design, most of  the existing  VLPC literatures \cite{Keskin_2018_Localization,Keskin_2019_Optimal}  mainly focus on  optimizing the resource allocation in different frequency bands to guarantee quality of service (QoS) of communication  and positioning.
 However, some  fundamental issues have not been well investigated.
Particularly,   does  the positioning benefit or compromise the communication?
How  are the two performances related? Given the limited power consumption, how to balance the two performances while taking the positioning error into account?

In this paper, we aim to address
the above mentioned fundamental issues, as well as to provide a robust beamforming and power allocation scheme.
The main contributions of this paper are
summarized as follows:

\begin{itemize}

\item  We establish a VLPC system  model with   a single   LED-lamp and  a mobile user with multiple photoelectric detectors (PD).
    By leveraging the fact that the VLC channel model  is a function of the receiver's location,  the lamp    estimate the channel state information (CSI)   based on the positioning results, instead of  transmitting pilot sequences for  CSI estimation, which can significantly reduce the system overhead.

\item  We derive  the  Cramer-Rao lower bound (CRLB) on the positioning error variance, which is used as the VLP  performance metric.
    In addition,
    we  derive the achievable rate expression for on-off keying (OOK) modulation, and its closed-form lower bound. Furthermore, by exploiting CRLB and achievable rate expressions,
    we  reveal the inner relationship between VLP and VLC for the first time, i.e.,
      derive the distribution of the CSI error of VLC based on the positioning error of VLP, and obtain a rate outage probability of VLC.

 \item  Based on the derived model and metrics, we further investigate a joint positioning and communication power allocation and beamforming problem to
{minimize the CRLB subject to rate outage constraints and power constraints.}
            The outage probability constraint  makes the optimization problem non-convex.
    Then, we apply the Conditional Value-at-Risk (CVaR) and the
    block coordinate descent (BCD) method
    techniques to convert the original problem into two  convex VLP and VLC sub-problems. Finally, we develop a BCD algorithm for robust VLPC design, in which the positioning and communication power allocation are iteratively optimized until convergence.

\end{itemize}

The rest of this paper is organized as follows.
We present the VLPC system model in Section \ref{section_ii}.
The key performance metrics for the VLPC system are derived in Section \ref{section_iii}.
In Section \ref{section_iv}, we investigate the chance constrained robust  design.
Extensive simulation results are presented in  Section \ref{section_v}. Section \ref{section_vi} concludes the paper. Moreover, Table \ref{Tab:notations} and \ref{Tab:acronyms}   present  the means of the  key notations and   the  main acronyms of  this paper, respectively.

\begin{table}[htpb]
	\caption{Summary of Key Notations}
	\label{Tab:notations}
	\centering
	\begin{tabular}{|c|l|}
		\hline
		\rule{0pt}{8pt}Notation  & Description   \\ \hline
		\rule{0pt}{7.5pt}${P_{\text{p}}}$ &  \tabincell{c}{Allocated positioning power} \\ \hline
		\rule{0pt}{7.5pt}$ {P_{\text{c}}}$ &  \tabincell{c}{Allocated communication power} \\ \hline
		\rule{0pt}{7.5pt}$ {{\bf{u}}}$ &
\tabincell{c}{Location vector of MU} \\ \hline
		\rule{0pt}{7.5pt}$ {{\bf{e}}_{\rm{p}}}$ &
\tabincell{c}{Positioning error vector} \\ \hline
		\rule{0pt}{7.5pt}$ {\bf{\hat h}}$ &  \tabincell{c}{Estimated CSI vector} \\ \hline
		\rule{0pt}{7.5pt}$ \Delta {\bf{h}}$ &
\tabincell{c}{CSI estimation error vector} \\ \hline
		\rule{0pt}{7.5pt}$ {\bf{I}}$ &
\tabincell{c}{Identity matrix } \\ \hline
		\rule{0pt}{7.5pt}$ {{\bf{J}}_{\rm{p}}}$ &
\tabincell{c}{Fisher information matrix} \\ \hline
		\rule{0pt}{7.5pt}$ R_{\rm{c}}^{\rm{L}}$ &
\tabincell{c}{Lower bound on the the achievable rate} \\ \hline
		\rule{0pt}{7.5pt}$\mathcal{P}$ &
\tabincell{c}{The set of distributions for ${\Delta {\bf{h}}}$} \\ \hline
		\rule{0pt}{7.5pt}$ \bar r$ &
\tabincell{c}{Minimum rate requirement} \\ \hline
		\rule{0pt}{7.5pt}$ {P_{{\rm{out}}}}$ &
\tabincell{c}{Maximum tolerable outage probability} \\ \hline
	\end{tabular}
\end{table}

\begin{table}[htpb]
	\caption{Summary of Main Acronyms}
	\label{Tab:acronyms}
	\centering
	\begin{tabular}{|c|l|}
		\hline
		\rule{0pt}{8pt}Notation  & Description   \\ \hline
		\rule{0pt}{7.5pt} VLP &  \tabincell{c}
{Visible light positioning} \\ \hline
		\rule{0pt}{7.5pt} VLC &  \tabincell{c}
{Visible light communication} \\ \hline
		\rule{0pt}{7.5pt} VLPC &  \tabincell{c}
{Visible light positioning and communication} \\ \hline
		\rule{0pt}{7.5pt} MU &  \tabincell{c}
{Mobile user} \\ \hline
        \rule{0pt}{7.5pt} CRLB &  \tabincell{c}
{Cramer-Rao lower bound} \\ \hline
        \rule{0pt}{7.5pt} CSI &  \tabincell{c}
{Channel state information} \\ \hline
        \rule{0pt}{7.5pt} BCD &  \tabincell{c}
{Block coordinate descent } \\ \hline
        \rule{0pt}{7.5pt} CVaR &  \tabincell{c}
{Conditional Value-at-Risk } \\ \hline
	\end{tabular}
\end{table}

$\textsl{Notations}$:
Boldfaced lowercase and uppercase letters represent
vectors and matrices, respectively.
$\mathcal{M}  \triangleq  \{ 1,...,M\}$.
The transpose  and trace of a matrix are denoted as
${\left(  \cdot  \right)^{\text{T}}}$  and
${\text{Tr}}\left(  \cdot  \right)$, respectively.
${\left\|  \cdot  \right\|_2}$ denotes 2-norm.
$\mathcal{N}$ denotes the Gaussian distribution.
$\bf{0}$  denotes a column vector where all elements are $0$.
$\mathbb{R}^{n}$  represents the space of $n$-dimensional real  matrices.
$\mathbb{S}^{n}$  represents the space of $n$-dimensional real symmetric matrices.

\section{System Model} \label{section_ii}

\begin{figure}[h]
      \centering
	\includegraphics[width=0.3\textwidth]{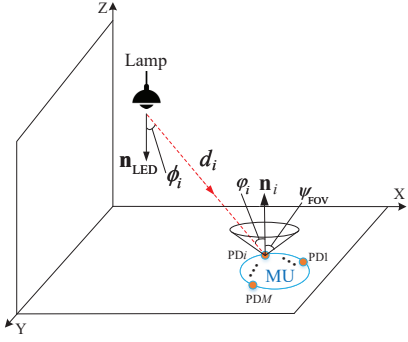}
    \caption{ System model illustration.}
  \label{model}
\end{figure}

{Considering a VLPC   system, as shown in Fig. \ref{model}, }
 where the   lamp is  equipped with  a single LED that points straight downward,
 and  a mobile user (MU) has a receiver with  $M$ PDs   ($M \ge 3$)\footnote{For 3D positioning, the number of   PDs  is at least 3.}.
Let ${\bf{l}} = {\left[ {{{x_l}},{y_l},{z_l}} \right]^{\text{T}}}$,
${{\bf{u}}} = {\left[ {{x_u},{y_u},{z_u}} \right]^{\text{T}}}$
and ${{\bf{u}}_i} = {\left[ {{x_i},{y_i},{z_i}} \right]^{\text{T}}}$ denote the locations of the LED, the MU and the $i$th PD, respectively, where $i \in \mathcal{M} $.
Moreover,  let  ${{\mathbf{v}}_i} = {\left[ {{v_{x,i}},{v_{y,i}},{v_{z,i}}} \right]^{\text{T}}}$ denote  the offset of the $i$th PD to the MU,  i.e., ${{\mathbf{u}}_i} = {\mathbf{u}} + {{\mathbf{v}}_i}$.

The wireless channel of the VLPC system has two types of links, i.e. the line-of-sight (LOS) link and the non line-of-sight (NLOS) link.
{ Generally,  the influence of
the  LOS  link is much stronger than that of the
NLOS link \cite{2004Komine}.}
In order to facilitate the theoretical analysis,
the design of VLPC system is based only on the  LOS link, but both the LOS link and NLOS link are considered in simulation verification.
According to the Lambert radiation model \cite{Wireless1997Kahn},  the LOS path gain  between the LED and the $i$th PD   within field-of-view (FoV)  can be expressed as
\begin{align}\label{h_LOS_1}
{h_{i}} = & \frac{{\left( {m+1} \right){A_{{\rm{PD}}}}}}{{2\pi d_i^2}}{\cos ^m}\left( {{\phi _{i}}} \right)\cos \left( {{\varphi _{i}}} \right)g{T_f}.
\end{align}
Here,  $m$ is the order of Lambertian emission and
$m =  - \frac{{\ln 2}}{{\ln \left( {\cos {\theta _{1/2}}} \right)}}$,  where ${{\theta _{1/2}}}$ is the semi-angle at half power.
Other parameters are defined as follows:
${A_{\text{PD}}}$ denotes the PD area;
$d_i$ is the distance between the LED and the $i$th PD;
${\phi _{i}}$ and ${\varphi _{i}}$ are the radiance  and incidence angles, respectively;
$g$ denotes the gain of the optical concentrator, and is given by
$g = \frac{{n_r^2}}{{{{\sin }^2}\left( {{\psi _{\text{FoV}}}} \right)}}$, where
$n_r$  denotes the refractive index, and
${\psi _{\text{FoV}}}$  represents the FoV of receiver; and
${T_f}$ denotes the  gain  of the optical filter.

Without loss of generality, assume that the PDs are pointing straight upward \cite{Xu_IPJ_2016}.
Based on the geometric relationship,  the LOS path gain \eqref{h_LOS_1} parameters can be specified as
 \begin{subequations}
\begin{align}
&{d_i} = {\left\| {{\bf{l}} - {{\bf{u}}_i}} \right\|_2},\\
&\cos \left( {{\phi _{i}}} \right) = \frac{{{{\left( {{{\bf{u}}_i} - {\bf{l}}} \right)}^{\text{T}}}{{\bf{n}}_{{\rm{LED}}}}}}{{{{\left\| {{\bf{l}} - {{\bf{u}}_i}} \right\|}_2}}} = \frac{{{z_l} - {z_i}}}{{{{\left\| {{\bf{l}} - {{\bf{u}}_i}} \right\|}_2}}},\\
&\cos \left( {{\varphi _{i}}} \right) = \frac{{{{\left( {{\bf{l}} - {{\bf{u}}_i}} \right)}^{\text{T}}}{{\bf{n}}_i}}}{{{{\left\| {{\bf{l}} - {{\bf{u}}_i}} \right\|}_2}}}= \frac{{{z_l} - {z_i}}}{{{{\left\| {{\bf{l}} - {{\bf{u}}_i}} \right\|}_2}}},
\end{align}
\end{subequations}
where ${{\bf{n}}_{{\rm{LED}}}}= {\left[ {0,0, - 1} \right]^{\rm{T}}}$ and
${{\bf{n}}_{i}}={\left[ {0,0, 1} \right]^{\rm{T}}}$ are unit direction vectors of the LED and the $i$th PD, respectively.
After substituting the above equations into \eqref{h_LOS_1},
the LOS path gain can be expressed as
\begin{align}\label{h_LOS_2}
{h_{i}} = \frac{{\alpha {{\left( {{z_l} - {z_i}} \right)}^{m + 1}}}}{{\left\| {{\bf{l}} - {{\bf{u}}_i}} \right\|_2^{m{\rm{ + 3}}}}},
\end{align}
where $\alpha  = \frac{{\left( {m + 1} \right){A_{{\rm{PD}}}}g{T_f}}}{{2\pi }}$.

\begin{figure}[h]
      \centering
	\includegraphics[width=0.3\textwidth]{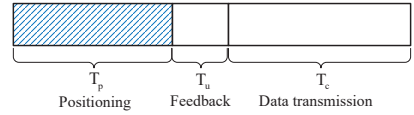}
    \caption{The frame structure of the considered VLPC system.}
  \label{frame}
\end{figure}

As shown in Fig. \ref{frame}, the operational frame consists of three subframes:
positioning subframe (downlink),
feedback subframe (uplink) and
data transmission  subframe (downlink).
The corresponding durations  are ${T_{\rm{p}}}$, ${T_{\rm{u}}}$ and ${T_{\rm{c}}}$, respectively.
More specific,   during the positioning  subframe, the LED lamp transmits the positioning symbols to the MU,
which estimates the PDs' locations  based on the RSS,
and the corresponding positioning information will be used for channel estimation in the next sub-frame.
Then, during the feedback  subframe, the MU sends feedback signals of the PDs' locations to   the lamp which  estimates the CSI between the LED and the PDs.
Finally, during  the data transmission subframe,   the lamp transmits data symbols to the   MU  according to the estimated CSI.
{This model can be extented into the multi-user system though proper multiple access methods. Due to the positioning theory, the positioning signal and subframe can be shared directly by all users without the multiple access. However, the multiple acess method, such as OFDMA, time division multiple address (TDMA), is necessary for the multi-user uplink feedback and downlink data transmission. Thus, The problem and solution also should refer to the classical theory of the multi-user networks.}

\subsection{Positioning Signal Model and  Measurements}

In the following, we will specify the signal model in order to analyze the operation in each subframe.
Let ${s_{{\rm{p}}}}\left( t \right)$ denote the positioning symbol generated at the lamp at time t, and
$\left| {{s_{\rm{p}}}}\left( t \right) \right| \le A$,
${\Bbb E} \left\{ {{s_{\rm{p}}}}\left( t \right) \right\} = 0$,
${\Bbb E}\left\{ {s_{\rm{p}}^2}\left( t \right) \right\} = \varepsilon $,
where $A >0 $ is the peak amplitude.

For $t \in \left[ {0,{T_{\rm{p}}}} \right]$,
the transmitted positioning signal $x_{\text{p}}\left( t \right)$ of the LED is given as
\begin{align}
{x_{\text{p}}}\left( t \right) = \sqrt {{P_{\text{p}}}} {s_{{\rm{p}}}}\left( t \right) + {I_{{\text{DC}}}},
\end{align}
where ${P_{\text{p}}}$ indicates the allocated transmission power to the positioning symbol, and ${I_{{\text{DC}}}}>0 $ denotes the direct current (DC) bias.

To guarantee that the transmitted signal is non-negative,
the power $P_{\rm{p}}$ should satisfy
\begin{align}\label{constraint_Pp_1}
\sqrt {{P_{\rm{p}}}} A \le {I_{{\rm{DC}}}}.
\end{align}

Given the  human eye safety  requirement,
 the LED optical power is limited,
i.e.$\sqrt {{P_{\rm{p}}}} A + {I_{{\rm{DC}}}} \le P_{\mathrm{o}}^{\max }$,
where $P_{\mathrm{o}}^{\max }$ denotes the maximum optical power.
Thus, the  power level $P_{\rm{p}}$ should also satisfy
\begin{align}\label{constraint_Pp_2}
\sqrt {{P_{\rm{p}}}}  \le \frac{{P_{\mathrm{o}}^{\max } - {I_{{\rm{DC}}}}}}{A}.
\end{align}

Besides, due to practical circuit limitations, the electrical power of the transmitted signal is  constrained as $\mathbb{E}\left\{ {x_{\text{p}}^2}\left( t \right) \right\} \le P_{\mathrm{e}}^{\max }$,  i.e.,
\begin{align}\label{constraint_Pp_3}
{P_{\rm{p}}}\varepsilon  + I_{{\rm{DC}}}^2 \le P_{\mathrm{e}}^{\max },
\end{align}
where $P_{\mathrm{e}}^{\max }$ denotes the maximum  LED  electrical  power.

{
Based on \eqref{constraint_Pp_1}, \eqref{constraint_Pp_2} and \eqref{constraint_Pp_3}, the constraint of the power $P_{\rm{p}}$ can be written as}
\begin{align}
0 \le {P_{\text{p}}} \le \min \left\{ \frac{{I_{{\rm{DC}}}^2}}{{{A^2}}},\frac{{{{\left( {P_{\mathrm{o}}^{\max } - {I_{{\rm{DC}}}}} \right)}^2}}}{{{A^2}}}, \frac{P_{\mathrm{e}}^{\max } - I_{{\rm{DC}}}^2}{\varepsilon}\right\}.
\end{align}

Then, the received positioning signal  at the $i$th PD can be expressed as
\begin{align}\label{received_position_signal}
{y_{{\rm{p}},i}}\left( t \right) = {h_i}{x_{\rm{p}}}\left( t \right) + {n_{{\rm{p}},i}}\left( t \right),
\end{align}
where  $n_{{\text{p}},i}$ denotes the received additive white Gaussian noise (AWGN), which includes shot noise and thermal noise \cite{Simulation_2017_Sindhubala},
and ${n_{{\rm{p}},i}} \sim {\cal N}\left( {{{0}},\sigma _{\rm{p}}^2} \right)$.

Theoretically, the electrical  power of the received positioning signal is given by
\begin{align}\label{receive_Pe}
{P_{{\rm{r}},i}} = {\mathbb{E}}\left\{ {y_{{\text{p}},i}^2}\left( t \right) \right\} =\left( {{P_{\rm{p}}}\varepsilon  + I_{{\rm{DC}}}^2} \right){h_{i}^2 }+ \sigma_{{\text{p}},i}^2,
\end{align}
where $\sigma_{{\text{p}},i}^2$ denotes noise  power.
Combining \eqref{h_LOS_2} and \eqref{receive_Pe}, we have the following $M$ equations
 \begin{align}\label{fangcheng}
\left\{ \begin{array}{l}
\frac{{{{\left( {{z_l} - {z_u} - {v_{z,1}}} \right)}^{m + 1}}}}{{\left\| {{\bf{l}} - {\bf{u}} - {{\bf{v}}_1}} \right\|_2^{m{\rm{ + 3}}}}} = \frac{1}{\alpha }{\left( {\frac{{{P_{{\rm{r}},1}} - \sigma _{{\rm{p}},1}^2}}{{{P_{\rm{p}}}\varepsilon + I_{{\rm{DC}}}^2}}} \right)^{\frac{1}{2}}},\\
{\rm{         }} \vdots \\
\frac{{{{\left( {{z_l} - {z_u} - {v_{z,M}}} \right)}^{m + 1}}}}{{\left\| {{\bf{l}} - {\bf{u}} - {{\bf{v}}_M}} \right\|_2^{m{\rm{ + 3}}}}} = \frac{1}{\alpha }{\left( {\frac{{{P_{{\rm{r}},M}} - \sigma _{{\rm{p}},M}^2}}{{{P_{\rm{p}}}\varepsilon  + I_{{\rm{DC}}}^2}}} \right)^{\frac{1}{2}}}.
\end{array} \right.
 \end{align}

To transform  the Equations in \eqref{fangcheng} into a concise form, we define the  auxiliary variable
\begin{align}
{\eta _i}\left( {\bf{u}} \right) = \frac{{{{\left( {{z_l} - {z_u} - {v_{z,i}}} \right)}^{m + 1}}}}{{\left\| {{\bf{l}} - {\bf{u}} - {{\bf{v}}_i}} \right\|_2^{m{\rm{ + 3}}}}} - \frac{1}{\alpha }{\left( {\frac{{{P_{{\rm{r}},i}} - \sigma _{{\rm{p}},i}^2}}{{{P_{\rm{p}}}\varepsilon + I_{{\rm{DC}}}^2}}} \right)^{\frac{1}{2}}}.
\end{align}
Thus, Equation \eqref{fangcheng} can be equivalently reformulated as  follows
\begin{align}
{\eta_i }\left( {\bf{u}} \right)= 0,i \in \mathcal{M}.\label{fangcheng_2}
\end{align}

Here  Equation \eqref{fangcheng_2} can be solved by   using off-the-shelf  optimization solvers, such as  FSOLVE in MATLAB\cite{matlab_fsolve}.

In general, the positioning error is inevitable.
Let   ${{{\mathbf{\hat u}}}}$ and ${{\bf{e}}_{\rm{p}}}$ denote  the estimated  MU location and the corresponding positioning error,
where ${{\bf{e}}_{\rm{p}}} = {\left[ {{e_x},{e_y},{e_z}} \right]^{\text{T}}}$.
Their relationship can be written as
\begin{align}\label{define_ep}
{{\mathbf{e}}_{\text{p}}} = {{\mathbf{u}}} - {{{\mathbf{\hat u}}}}.
\end{align}

Generally, the positioning error ${{\mathbf{e}}_{\mathrm{p}}}$ can be assumed to follow the  Gaussian distribution \cite{Wang2009,Lottici_JSAC_2002,Gezici_SPM_2005},
and then  CRLB  can be achieved  by the maximum-likelihood (ML)
estimator \cite{Kay_1993,Cheung_2004_Least}.
We use  $f_{\mathbf{e}_{\mathrm{p}}}\left( \mathbf{e}_{\rm{p}} \right)$ to denote the   probability density distribution of ${{\bf{e}}_{\rm{p}}}$, which follows  a Gaussian distribution with  mean ${\bf{0}}$   and  covariance matrix  ${{\bf{E}}_{\rm{p}}}$,
i.e.,
${{\bf{e}}_{\rm{p}}} \sim {\mathcal{N}}\left( {{\bf{0}},{{\bf{E}}_{\rm{p}}}} \right)$.

\subsection{MU Feedback and Channel Estimation}

When $t \in \left[ {{T_{\rm{p}}}, {T_{\rm{p}}}+{T_{\rm{u}}}} \right]$, the estimated location of the MU ${{{\mathbf{\hat u}}}}$ will be  sent to the lamp, which also serves as an anchor node.
Based on ${{{\mathbf{\hat u}}}}$, the lamp can estimate the CSI between the LED and the PDs.
 Specifically, ${\bf{\hat h}} = \left[ {{{\hat h}_1},...,{{\hat h}_M}} \right]^{\mathrm{T}} \in {\mathbb{R}^{M \times 1}}$ denotes the estimated CSI vector;
$\Delta {\bf{h}} = \left[ {\Delta {h_1},...,\Delta {h_M}} \right]^{\mathrm{T}} \in {\mathbb{R}^{M \times 1}}$  denotes the CSI estimation error vector.

 Let ${h_{i}}$ and ${\hat h_{i}}$ denote  the perfect  and estimated CSI between the LED and the $i$th PD,
and $\Delta { h_{i}}$ denote  the estimated CSI error, i.e., ${h_{i}}={\hat h_{i}}+\Delta { h_{i}}$.
According to \eqref{h_LOS_2}, the estimated CSI ${\hat h_{i}}$  is {
a function of the estimated UE's location  ${{{\mathbf{\hat u}}}}$  given by }
\begin{align}\label{h_hat}
{\hat h_{i}} = \frac{{\alpha {{\left( {{z_l} - {{\hat z}_u} - {v_{z,i}}} \right)}^{m + 1}}}}{{\left\| {{\bf{l}} - {\bf{\hat u}} - {{\bf{v}}_i}} \right\|_2^{m + 3}}}.
\end{align}
Based on \eqref{h_LOS_2} and \eqref{h_hat},  the estimated CSI error $\Delta { h_{i}}$ is given as
\begin{align}\label{Delta h}
\Delta {h_i}{\rm{ }} = \frac{{\alpha {{\left( {{z_l} - {{\hat z}_u} - {v_{z,i}} -{e_z}} \right)}^{m + 1}}}}{{\left\| {{\bf{l}} - {\bf{\hat u}} - {{\bf{v}}_i} - {{\bf{e}}_{\rm{p}}}} \right\|_2^{m + 3}}} - \frac{{\alpha {{\left( {{z_l} - {{\hat z}_u} - {v_{z,i}}} \right)}^{m + 1}}}}{{\left\| {{\bf{l}} - {\bf{\hat u}} - {{\bf{v}}_i}} \right\|_2^{m + 3}}}.
\end{align}

\subsection{Data Transmission}

Let ${s_{{\rm{OOK}}}}\left( t \right)$ denote the  data symbol transmitted from the LED, and ${s_{{\rm{OOK}}}}\left( t \right)$  takes value $0$ or $A$ with equal probability, i.e.,
$\Pr \left\{ {{s_{{\rm{OOK}}}}\left( t \right) = 0} \right\} = \frac{1}{2}$, and
${\Pr \left\{ {{s_{{\rm{OOK}}}}\left( t \right) = A} \right\} = \frac{1}{2}}$,
where $A$ is the peak amplitude of the symbol. Due to $s_{\mathrm{OOK}}\geq 0$, $I_{\mathrm{DC}}$ can be $0$ in the data transmission.

For $t \in \left[ {{T_{\rm{p}}}+{T_{\rm{u}}}, {T_{\rm{p}}}+{T_{\rm{u}}}+{T_{\rm{c}}}} \right]$,
the LED transmitted data signal $x_{\rm{c}}\left( t \right)$ can be expressed as
\begin{align}
{x_{\text{c}}}\left( t \right) = \sqrt {{P_{\text{c}}}} {s_{{\text{OOK}}}}\left( t \right),
\end{align}
where ${{P_{\text{c}}}}$ indicates the allocated communication power of the LED.

Similarly, the communication power ${{P_{\text{c}}}}$  should also meet the eye safety constraint, i.e.
$\sqrt {{P_{\rm{c}}}} A \le P_{\mathrm{o}}^{\max }$, where $P_{\mathrm{o}}^{\max }$ denotes the maximum optical power.
Thus, the communication power $P_{\rm{c}}$ should satisfy
\begin{align}\label{constraint_Pc_1}
\sqrt {{P_{\rm{c}}}}  \le \frac{{P_{\mathrm{o}}^{\max }}}{A}.
\end{align}

{
Under practical circuit limitations, the electrical power of the transmitted signal is  constrained as $\mathbb{E}\left\{ {x_{\text{c}}^2\left( t \right)} \right\} \le P_{\mathrm{e}}^{\max }$,  i.e.,}
\begin{align}\label{constraint_Pc_2}
{P_{\rm{c}}}\mathbb{E}\left\{ {s_{{\rm{OOK}}}^2}\left( t \right) \right\} = \frac{P_{\mathrm{c}}A^2}{2} \le P_{\mathrm{e}}^{\max },
\end{align}
where $P_{\mathrm{e}}^{\max }$ denotes the maximum  LED electrical  power.

Based on \eqref{constraint_Pc_1} and \eqref{constraint_Pc_2}, the power $P_{\rm{c}}$ should satisfy
\begin{align}
0 \le {P_{\text{c}}} \le  \min \left\{ {\frac{{{{\left( {P_{\mathrm{o}}^{\max }} \right)}^2}}}{{{A^2}}},\frac{{2P_{\mathrm{e}}^{\max }}}{{{A^2}}}} \right\}.
\end{align}

At the receiver, let  ${\mathbf{v}} = \left[ {{v_1},...,{v_M}} \right]^{\text{T}} \in {\mathbb{R}^{M \times 1}}$ denote the receive beamforming vector of the MU, and $\left\| {\bf{v}} \right\| = 1$.
Therefore, the received data signal at MU can be expressed as
\begin{align}\label{received_data_signal}
{y_{\text{c}}}\left( t \right) = {{\mathbf{v}}^{\text{T}}}{\left( {{\bf{\hat h}} + \Delta {\bf{h}}} \right)}{x_{\text{c}}}\left( t \right) + z_{\text{c}},
\end{align}
where
$z_{\text{c}} \buildrel \Delta \over = {{\mathbf{v}}^{\text{T}}}{{\mathbf{n}}_{\text{c}}}$, and
${{\mathbf{n}}_{\text{c}}} \in {\mathbb{R}^{M\times 1}}$ denotes the receiver Gaussian noise vector,    i.e.,
${{\bf{n}}_{\bf{c}}} \sim {\cal N}\left( {{\bf{0}},\sigma _{\rm{c}}^2{\bf{I}}} \right)$.

\section{Performance Metrics} \label{section_iii}

\subsection{Cramer-Rao Lower Bound }

The  CRLB represents a lower bound on the variance of the positioning estimation error. Hence,
we adopt CRLB as the performance metric for the positioning accuracy in this paper.
Specifically, we consider  three-dimensional MU location estimation.
Considering the received signal model in \eqref{received_position_signal},
the likelihood function of  ${{y_{{\rm{p}},i}}}\left( t \right)$ can be written as
\begin{align}
f\left( {{y_{{\text{p}},i}}\left( t \right);{{\mathbf{u}}}} \right) = \frac{1}{{\sqrt {2\pi } {\sigma _{\rm{p}}}}}{e^{ - \frac{{{{\left( {{y_{{\rm{p}},i}}\left( t \right) - {h_i}{x_{\rm{p}}}\left( t \right)} \right)}^2}}}{{2\sigma _{\rm{p}}^2}}}} .
\end{align}
Therefore, the log-likelihood function of the received signal $\left\{ {{y_{{\rm{p}},i}}}\left( t \right) \right\}_i^M$ is obtained as follows \cite{Keskin2018Direct}
\begin{align}\label{log_likehood}
\Lambda \left( {{{\mathbf{u}}}} \right) &
=\ln \left( \mathop \prod \limits_{i = 1}^M f\left( {{y_{{\text{p}},i}}\left( t \right);{{\mathbf{u}}}} \right)  \right)\nonumber\\
&= \ln \kappa - \frac{1}{{2{\sigma _{\rm{p}}^2}}}\sum\limits_{i = 1}^M {\int_0^{T_{\rm{p}}} {{{\left( {{y_{{\text{p}},i}}\left( t \right) - {h_{i}}{x_{\text{p}}}\left( t \right)} \right)}^2}\,\mathrm{d}t} },
\end{align}
where
$\kappa$ is a constant that does not depend on the unknown parameters.
Recalling the definition given in \eqref{define_ep}, and denoting by ${{\bf{E}}_{\rm{p}}}$  the covariance matrix of the positioning error ${{{\bf{e}}_{\rm{p}}}}$. Then, according to the definition of  the CRLB  on the variance of any unbiased estimator \cite{Gander1990An},
a lower limit on the variance of the $i$th element  in an unbiased estimate vector ${\mathbf{\hat u}}$  is given by
\begin{align}
\sum_{i}{\left[ {{{\bf{E}}_{\rm{p}}}} \right]_{ii}} \ge \sum_{i}{\left[ {{\bf{J}}_{\rm{p}}^{ - 1}} \right]_{ii}},
\end{align}
where
${\left[ \cdot  \right]_{ii}}$ denotes  the diagonal element of a matrix,  and
${\bf{J}}_{\rm{p}}$ denotes the Fisher Information  matrix (FIM),
which is defined as
\begin{align}
{\left[ {{{\mathbf{J}}_{\text{p}}}} \right]_{ij}} =
 - {\mathbb{E}}\left\{ {\frac{{{\partial ^2}\Lambda \left( {\mathbf{u}} \right)}}
{{\partial {{\mathbf{u}}_i}\partial {{\mathbf{u}}_j}}}} \right\},
\end{align}
for $i \in \left\{1,2,3\right\}$, $j \in \left\{1,2,3\right\}$. Likewise, ${\left[ \cdot  \right]_{ij}}$ denotes the element on the $i$th row and $j$th column of a matrix.
We show in Appendix  \ref{tuidao_FIM} that the FIM for \eqref{log_likehood} is given by
\begin{align}\label{FIM}
{\bf{J}}_{\rm{p}} & = \frac{{{T_{\rm{p}}}{\left( {P_{\rm{p}}}\varepsilon + I_{{\rm{DC}}}^2\right)}}}{{\sigma _{\rm{p}}^2}} {\bf{Q}},
\end{align}
where
\begin{subequations}
\begin{align}
&{\bf{Q}}= \left[ {\begin{array}{*{20}{c}}
{\sum\limits_{i = 1}^M {\frac{{\partial {h_i}}}{{\partial {x_u}}}\frac{{\partial {h_i}}}{{\partial {x_u}}}} }&{\sum\limits_{i = 1}^M {\frac{{\partial {h_i}}}{{\partial {x_u}}}\frac{{\partial {h_i}}}{{\partial {y_u}}}} }&{\sum\limits_{i = 1}^M {\frac{{\partial {h_i}}}{{\partial {x_u}}}\frac{{\partial {h_i}}}{{\partial {z_u}}}} }\\
{\sum\limits_{i = 1}^M {\frac{{\partial {h_i}}}{{\partial {x_u}}}\frac{{\partial {h_i}}}{{\partial {y_u}}}} }&{\sum\limits_{i = 1}^M {\frac{{\partial {h_i}}}{{\partial {y_u}}}\frac{{\partial {h_i}}}{{\partial {y_u}}}} }&{\sum\limits_{i = 1}^M {\frac{{\partial {h_i}}}{{\partial {y_u}}}\frac{{\partial {h_i}}}{{\partial {z_u}}}} }\\
{\sum\limits_{i = 1}^M {\frac{{\partial {h_i}}}{{\partial {x_u}}}\frac{{\partial {h_i}}}{{\partial {z_u}}}} }&{\sum\limits_{i = 1}^M {\frac{{\partial {h_i}}}{{\partial {y_u}}}\frac{{\partial {h_i}}}{{\partial {z_u}}}} }&{\sum\limits_{i = 1}^M {\frac{{\partial {h_i}}}{{\partial {z_u}}}\frac{{\partial {h_i}}}{{\partial {z_u}}}} }
\end{array}} \right],\\
&\frac{{\partial {h_i}}}{{\partial {x_u}}} = \frac{{ - \alpha \left( {m + 3} \right){{\left( {{z_l} - {z_u} - {v_{z,i}}} \right)}^{m + 1}}\left( {{x_u} + {v_{x,i}} - {x_l}} \right)}}{{{{\left\| {{\bf{l}} - {\bf{u}} - {{\bf{v}}_i}} \right\|}^{m + 5}}}},  \\
&\frac{{\partial {h_i}}}{{\partial {y_u}}} = \frac{{ - \alpha \left( {m + 3} \right){{\left( {{z_l} - {z_u} - {v_{z,i}}} \right)}^{m + 1}}\left( {{y_u} + {v_{y,i}} - {y_l}} \right)}}{{{{\left\| {{\bf{l}} - {\bf{u}} - {{\bf{v}}_i}} \right\|}^{m + 5}}}},\\
&\frac{{\partial {h_i}}}{{\partial {z_u}}} = \frac{{ - \left( {m + 1} \right)\alpha {{\left( {{z_l} - {z_u} - {v_{z,i}}} \right)}^m}}}{{{{\left\| {{\bf{l}} - {\bf{u}} - {{\bf{v}}_i}} \right\|}^{m + 3}}}} \nonumber\\
&\quad\quad\quad\quad\quad\quad + \frac{{\left( {m + 3} \right)\alpha {{\left( {{z_l} - {z_u} - {v_{z,i}}} \right)}^{m + 2}}}}{{{{\left\| {{\bf{l}} - {\bf{u}} - {{\bf{v}}_i}} \right\|}^{m + 5}}}}.
\end{align}
\end{subequations}

Moreover, let $B$ (Hz) denote  the bandwidth of VLC link.
Combine the bandwidth of VLC link,
the variance of positioning  error ${{\bf{e}}_{\rm{p}}}$  is  lower bounded by
\begin{align}\label{CRLB}
    {\text{Tr}}\left( {{{\mathbf{E}}_{\text{p}}}} \right) \geq {\text{Tr}}\left( {{\mathbf{J}}_{\text{p}}^{ - 1}} \right) = \frac{{B\sigma _{\text{p}}^2{\text{Tr}}\left( {{{\mathbf{Q}}^{ - 1}}} \right)}}
    {{{T_{\text{p}}}\left( {{P_{\text{p}}}\varepsilon + I_{{\text{DC}}}^2} \right)}}.
\end{align}

In the following, we will use \eqref{CRLB} as the positioning performance metric  which is to be minimized.

\subsection{Achievable Rate for OOK}

When considering  OOK modulation, the input signal  no longer follows the Gaussian distribution, and thus the Shannon capacity formula based on the Gaussian assumption cannot be directly applied.
In Appendix \ref{tuidao_Rc}, we show that the mutual information is given by
 \begin{align}\label{Rate}
&I\left( {{x_{\text{c}}};{y_{{\text{c}}}}} \right)
= - \frac{1}{2}{\mathbb{E}_{{z_{\rm{c}}}}}\left\{ {{{\log }_2}\left( {\frac{{{e^{ - \frac{{z_{\rm{c}}^2}}{{2\sigma _{\rm{c}}^2}}}} + {e^{ - \frac{{{{\left( {{{\bf{v}}^T}{\bf{h}}\sqrt {{P_{\rm{c}}}} A + {z_{\rm{c}}}} \right)}^2}}}{{2\sigma _{\rm{c}}^2}}}}}}{2}} \right)} \right\} \nonumber \\
& - \frac{1}{2}{\mathbb{E}_{{z_{\rm{c}}}}}\left\{ {{{\log }_2}\left( {\frac{{{e^{ - \frac{{z_{\rm{c}}^2}}{{2\sigma _{\rm{c}}^2}}}} + {e^{ - \frac{{{{\left( { - {{\bf{v}}^T}{\bf{h}} \sqrt {{P_{\rm{c}}}} A + {z_{\rm{c}}}} \right)}^2}}}{{2\sigma _{\rm{c}}^2}}}}}}{2}} \right)} \right\}- \frac{1}{{2\ln 2}}.
\end{align}

Due to the expectation operation, the  expression \eqref{Rate} is not analytically tractable, and can only be calculated numerically at the expense of high computational complexity. To strike a balance between complexity and analytical tractability, we derive a closed-form lower bound  on the  mutual information  \eqref{Rate}.

{
Let $R_{\rm{c}}^{\rm{L}}\left( {\Delta {\bf{h}}} \right)$ denote  a lower bound on the  achievable rate.}
Using Jensen's Inequality  and  combining  ${\bf{h}} = {\bf{\hat h}} + \Delta {\bf{h}}$,   we show in Appendix \ref{tuidao_Rc_Ub_Lb} that $R_{\rm{c}}^{\rm{L}}\left( {\Delta {\bf{h}}} \right)$ with the bandwidth $B$ is given by
\begin{align}\label{Rate_lowerbound}
R_{\rm{c}}^{\rm{L}}\left( {\Delta {\bf{h}}} \right) = 3B &- \frac{B}{{\ln 2}} \nonumber\\
&- 2B{\log _2}\left( {1 + {e^{ - \frac{{{{\left( {{{\bf{v}}^{\rm{T}}}\left( {{\bf{\hat h}} + \Delta {\bf{h}}} \right)} \right)}^2}{P_{\rm{c}}}{A^2}}}{{4B\sigma _{\rm{c}}^2}}}}} \right).
\end{align}

The CSI error $ \Delta {\bf{h}}$  affects the achievable rate, and ${\Delta {\bf{h}}}$ is  a function of the positioning error ${{{\mathbf{e}}_{\text{p}}}}$ as shown in Equation \eqref{Delta h},
which also depends on the positioning signal power ${P_{\text{p}}}$.
Therefore, both the positioning signal power ${P_{\text{p}}}$ and communication signal power ${P_{\text{c}}}$ affect the achievable  rate,  and their allocations need to be carefully optimized.

\section{Robust Power Allocation for VLPC Design} \label{section_iv}

In this section, we investigate the  positioning error variance minimization problem via a  robust power allocation design by considering the rate outage probability constraint.
Different from existing works,  this paper considers the application of the VLPC system in the 3D case, and establishes the connection between the rate outage probability and the positioning error  for the first time.

\subsection{Problem Formulation}

We derive the rate outage probability by investigating the relationship between the positioning error and CSI error.
Based on \eqref{Delta h}, the CSI error  ${\Delta {\bf{h}}}$ is  a function of the positioning error ${{{\mathbf{e}}_{\text{p}}}}$.
Thus, let  $\Delta {h_i}  \buildrel \Delta \over = g_i\left( {{\bf{e}}_{\rm{p}}} \right)$ denote the CSI error function of the $i$th PD,  i.e.,
\begin{align} \label{function Delta h}
\Delta {h_i}={g_i}\left( {{{\bf{e}}_{\rm{p}}}} \right)=&
\frac{{\alpha {{\left( {{z_l} - {{\hat z}_u} - {v_{z,i}} - {e_z}} \right)}^{m + 1}}}}{{\left\| {{\bf{l}} - {\bf{\hat u}} - {{\bf{v}}_i} - {{\bf{e}}_{\rm{p}}}} \right\|_2^{m + 3}}}\nonumber\\
&\quad\quad\quad\quad - \frac{{\alpha {{\left( {{z_l} - {{\hat z}_u} - {v_{z,i}}} \right)}^{m + 1}}}}{{\left\| {{\bf{l}} - {\bf{\hat u}} - {{\bf{v}}_i}} \right\|_2^{m + 3}}},
\end{align}
where $i \in \mathcal{M}$.

Then, we can write  ${{\bf{e}}_{\rm{p}}} = g_i^{ - 1}\left( {\Delta {h_i}} \right)$,
and  the probability density function
${f_{{h_i}}}\left( {\Delta {h_i}} \right)$ is given by
\begin{align}
 {f_{{h_i}}}\left( {\Delta {h_i}} \right) = {f_{\bf{e}_{\mathrm{p}}}}\left( {g_i^{ - 1}\left( {\Delta {h_i}} \right)} \right)\left| {\frac{{\partial g_i^{ - 1}\left( {\Delta {h_i}} \right)}}{{\partial \Delta {h_i}}}} \right|.
 \end{align}
Unfortunately, an explicit expression of the function ${{\bf{e}}_{\rm{p}}} = g_i^{ - 1}\left( {\Delta {h_i}} \right)$ is difficult to derive.

Nonetheless, we can numerically   calculate both the mean and covariance matrix of the   CSI error vector ${\Delta {\bf{h}}}$.
Specifically,  let
${\bm{\mu }} = \mathbb{E}\left\{ {\Delta {\mathbf{h}}} \right\}\in {\mathbb{R}^{M \times 1}}$ denote   the mean vector of the estimated CSI error   ${\Delta {\bf{h}}}$, which is given as
\begin{align}
 \mathbb{E}\left\{ {\Delta {{h_i}}} \right\}= \int {{g_i}\left( {{{\bf{e}}_{\rm{p}}}} \right){{f_{{{\bf{e}_{\mathrm{p}}}}}}\left( {{\bf{e}}_{\rm{p}}} \right)}\,\mathrm{d}{{\bf{e}}_{\rm{p}}}}.
\end{align}
Furthermore, let ${\bf{D}} = \mathbb{E}\left\{ {\left( {\Delta {\bf{h}} - {\bm{\mu }}} \right){{\left( {\Delta {\bf{h}} - {\bm{\mu }}} \right)}^T}} \right\}\in {\mathbb{R}^{M \times M}} $
denote the covariance matrix of the estimated CSI error vector ${\Delta {\bf{h}}}$.
Then, the element on the $i$th row and $j$th column of ${\bf{D}}$  is given by
\begin{align}\label{covariance matrix}
{\left[ {\mathbf{D}} \right]_{ij}}  =   \mathbb{E}\left\{\left( {\Delta {h_i} - \mathbb{E}\left\{ {\Delta {h_i}} \right\}} \right)\left( {\Delta {h_j} - \mathbb{E}\left\{ {\Delta {h_j}} \right\}} \right) \right\},
\end{align}
where $i \in \mathcal{M}$, and $j \in \mathcal{M}$.

The exact  distribution of CSI errors  ${\Delta {\bf{h}}}$ is    unknown except for its first and second-order moments.
Then, we may define a set $\mathcal{P}$ of   distributions for ${\Delta {\bf{h}}}$ as follows
\begin{align}\label{Delta_h_distribution}
 \mathcal{P} =
 \left\{{\mathbb{P}:{\mathbb{E}_\mathbb{P}}\left\{ {\Delta {\bf{h}}} \right\} = {\bm{\mu }}, {\rm{Var}_\mathbb{P}}\left\{ {\Delta {\bf{h}}} \right\}={\bf{D}}} \right\},
\end{align}
where $\mathbb{P}$ denotes an arbitrary distribution with the mean ${\bm{\mu }}$ and covariance matrix ${\bf{D}}$.
The  set $\mathcal{P}$ in  \eqref{Delta_h_distribution} determines the CSI error variation, and the rate outage probability.

Now, we can formulate the positioning error variance minimization through robust power allocation problem
as  follows
\begin{subequations}\label{tradeoff_originalPro1}
\begin{align}
\mathop {\min }\limits_{{P_{\rm{p}}},{P_{\rm{c}}},{\bf{v}}} &{\rm{Tr}}\left( {{\mathbf{J}}_{\text{p}}^{ - 1}} \right)\\
{\rm{s}}.{\rm{t}}.
&  \mathop {\sup }\limits_{\Delta {\bf{h}} \sim {\mathbb{P}},~\mathbb{P} \in {\cal P}} \Pr \left\{ {R_{\rm{c}}^{\rm{L}}\left( {\Delta {\bf{h}}} \right) \le \bar r} \right\} \le {P_{{\rm{out}}}},
\label{rate_yueshu}\\
& {P_{\rm{p}}} + {P_{\rm{c}}} \le {{P_{\rm{T}}}},\label{robust_cons_Ptotal}\\
& 0 \le {P_{\text{p}}} \le  P_{\rm{p}}^{\max}, \label{constraint_Pp}\\
& 0 \le {P_{\text{c}}} \le  P_{\rm{c}}^{\max}, \label{constraint_Pc}\\
& \left\| {\bf{v}} \right\|^2 = 1, \label{norm_v}
\end{align}
\end{subequations}
where
$\bar r$ denotes the minimum rate requirement,
${P_{{\rm{out}}}}$ denotes the maximum  tolerable outage probability,
${{P_{\rm{T}}}}$ denotes the total power,
$P_{\rm{p}}^{\max } \buildrel \Delta \over =  \min \left\{ \frac{{I_{{\rm{DC}}}^2}}{{{A^2}}},\frac{{{{\left( {P_{\mathrm{o}}^{\max } - {I_{{\rm{DC}}}}} \right)}^2}}}{{{A^2}}}, \frac{P_{\mathrm{e}}^{\max } - I_{{\rm{DC}}}^2}{\varepsilon}\right\}$, and
$P_{\rm{c}}^{\max } \buildrel \Delta \over =  \min \left\{ {\frac{{{{\left( {P_{\mathrm{o}}^{\max }} \right)}^2}}}{{{A^2}}},\frac{{2P_{\mathrm{e}}^{\max }}}{{{A^2}}}} \right\}$.

\subsection{Proposed Robust VLPC Method}

The main challenge of problem \eqref{tradeoff_originalPro1}  lies in the chance constraint \eqref{rate_yueshu}, which does not have a closed-form expression.
Hence, we will reformulate constraint \eqref{rate_yueshu}.
Combined with the lower bound on the achievable rate in \eqref{Rate_lowerbound},
the inequality ${R_{\rm{c}}^{\rm{L}}\left( {\Delta {\bf{h}}} \right) \le \bar r}$ can be equivalently rewritten as
\begin{align}\label{rate_yueshu_bian1}
{\left( {{{\bf{v}}^{\text{T}}}\left( {{\bf{\hat h}} + \Delta {\bf{h}}} \right)} \right)^2} \le  \frac{\delta }{{{P_{\rm{c}}}}},
\end{align}
where $\delta  \buildrel \Delta \over =   - \frac{{4B\sigma _{\rm{c}}^2}}{{{A^2}}}\ln \left( {{2^{\frac{3}{2} - \frac{1}{{2\ln 2}} - \frac{{\bar r}}{{2B}}}} - 1} \right)$.
By using the following equivalence relationship
\begin{align}
{\bf{V}} = {\bf{v}}{{\bf{v}}^{\text{T}}} \Leftrightarrow
{\bf{V}} \succeq {\bf{0}},
{\rm{rank}}\left( {\bf{V}} \right) = 1,
\end{align}
and neglecting the non-convex rank constraint ${\rm{rank}}\left( {\bf{V}} \right) = 1$,
constraints \eqref{rate_yueshu_bian1} and \eqref{norm_v}   can be  respectively relaxed as
\begin{subequations}
\begin{align}
&\Delta {{\bf{h}}^{\text{T}}}{\bf{V}}\Delta {\bf{h}} + 2{{{\bf{\hat h}}}^{\text{T}}}{\bf{V}}\Delta {\bf{h}} + {{{\bf{\hat h}}}^{\text{T}}}{\bf{V\hat h}} \le  \frac{\delta }{{{P_{\rm{c}}}}},\label{semidefinite V1}\\
&{\rm{Tr}}\left( {\bf{V}} \right) = 1,~{\bf{V}} \succeq {\bf{0}},\label{semidefinite V2}
\end{align}
\end{subequations}
In words, we exploit the semidefinite relaxation (SDR) technique to relax  \eqref{rate_yueshu_bian1} to a semidefinite program (SDP).
Then, the outage constraint \eqref{rate_yueshu} can be recast as
\begin{align}
\Pr \left\{ {\Delta {{\bf{h}}^{\text{T}}}{\bf{V}}\Delta {\bf{h}} + 2{{{\bf{\hat h}}}^{\text{T}}}{\bf{V}}\Delta {\bf{h}} + {{{\bf{\hat h}}}^{\text{T}}}{\bf{V\hat h}} - \frac{\delta }{{{P_{\rm{c}}}}} \le 0} \right\} \le {P_{{\rm{out}}}}.\label{rate_yueshu_bian2}
\end{align}

An effective approach to proceed is to transform \eqref{rate_yueshu_bian2} into  a distributionally robust chance constraint. Then, we can find the worst-case distribution
among all the possible distributions from the ambiguity
set, i.e.,
\begin{align}\label{CRaV1}
{\inf_{\mathbb{P} \in \mathcal{P}}} {\mathrm{Pr}_{\mathbb{P}}}& \left\{\Delta\mathbf{h}^\mathrm{T}(-\mathbf{V})\Delta\mathbf{h}+2\Delta\mathbf{h}^\mathrm{T}(-\mathbf{V})\hat{\mathbf{h}}+\hat{\mathbf{h}}^\mathrm{T}(-\mathbf{V})\hat{\mathbf{h}}\right.\nonumber\\
&\quad\quad\quad\quad\quad\quad\quad\quad\left.+\frac{\delta}{P_{\mathrm{c}}} \leq 0 \right\} \ge 1 - {P_{{\rm{out}}}},
\end{align}
where $\mathop {\inf }\limits_{\mathbb{P} \in \mathcal{P}} {\Pr _\mathbb{P}}\left\{  \cdot  \right\}$ denotes
the  distribution that can achieve the minimum value of the probability.

To further deal with the intractability of \eqref{CRaV1}, we introduce a CVaR-based method \cite{Steve2013Distributionally}, which is known as a good convex approximation of the worst-case chance constraint.

 $\textbf{Lemma} \ 1$ (CVaR-Based Method):
For a constraint function $L$ that is concave or quadratic in ${\bm{\xi }}$, the distributionally robust chance constraint is equivalent to the worst-case constraint,
given by \cite{Backscatter2019Zhang}
\begin{align}\label{P_CVaR}
\mathop {\inf }\limits_{\mathbb{P} \in \mathcal{P}}  {{\Pr}_\mathbb{P}}\left\{ {L\left( {\bm{\xi }} \right) \le 0} \right\} &\ge 1 - \rho \nonumber\\
\Leftrightarrow&
 \mathop {\sup }\limits_{\mathbb{P} \in \mathcal{P}} \left\{ \mathbb{P}-{{\text{CVaR}}_\rho }\left\{ {L\left( {\bm{\xi }} \right)} \right\} \right\} \le 0,
\end{align}
where the expression  $\mathbb{P}-{{\text{CVaR}}_\rho }\left\{ {L\left( {\bm{\xi }} \right)} \right\}$
denotes the CVaR of function $L\left( {\bm{\xi }} \right)$
at threshold $\rho$ under distribution $\mathbb{P}$,  which is defined as
\begin{align}
\mathbb{P} - {\text{CVa}}{{\text{R}}_\rho }\left\{ {L\left( {\bm{\xi }} \right)} \right\} = \mathop {\inf }\limits_{\beta  \in \mathbb{R}} \left\{ {\beta  + \frac{1}{\rho }{\mathbb{E}_\mathbb{P}}\left[ {{{\left( {L\left( {\bm{\xi }} \right) - \beta } \right)}^ + }} \right]} \right\}.
\end{align}
Here,  $\mathbb{R}$ is the set of real numbers, ${\left( z \right)^ + } = \max \left\{ {0,z} \right\}$, and $\beta \in \mathbb{R} $ is an auxiliary variable introduced by CVaR.
The worst-case CVaR on the right hand side of \eqref{P_CVaR} can be converted into a group of SDPs, which will be shown in the following lemma.

$\textbf{Lemma} \ 2$: Let $L\left( {\bm{\xi }} \right) = {{\bm{\xi }}^{\text{T}}}{\mathbf{Q}}{\bm{\xi }} + {{\mathbf{q}}^{\text{T}}}{\bm{\xi }} + {{\mathbf{q}}^0}$ denote a quadratic function of ${\bm{\xi }}$, $\forall {\bm{\xi }} \in {\mathbb{R}^n}$.
The worst-case CVaR can be computed  as \cite{Backscatter2019Zhang}
\begin{subequations}\label{eqn:lemma2}
\begin{align}
\mathop {\sup }\limits_{\mathbb{P} \in \mathcal{P}} &\left\{ \mathbb{P}-{{\text{CVaR}}_\rho }\left\{ {L\left( {\bm{\xi }} \right)} \right\} \right\}  =  \mathop {\min }\limits_{\beta ,{\mathbf{M}}} \left\{ \beta  + \frac{1}{\rho }{\text{Tr}}\left( {{\mathbf{\Omega M}}} \right)\right\}  \\
{\text{s}}.{\text{t}}.&{\mathbf{M}} \succeq {\mathbf{0}},{\mathbf{M}} \in {\mathbb{S}^{n + 1}},\\
&{\mathbf{M}} - \left[ {\begin{array}{*{20}{c}}
  {\mathbf{Q}}&{\frac{1}{2}{\mathbf{q}}} \\
  {\frac{1}{2}{\mathbf{q}}^\mathrm{T}}&{{{\mathbf{q}}^0} - \beta }
\end{array}} \right] \succeq {\mathbf{0}},
\end{align}
\end{subequations}
where ${\mathbf{M}}$ is an auxiliary  matrix variable, and ${\mathbf{\Omega }}$ is a matrix defined as
\begin{align}
{\mathbf{\Omega }} = \left[ {\begin{array}{*{20}{c}}
  {{\mathbf{\Sigma}} + {\bm{\mu}}{{\bm{\mu}}^{\text{T}}}}&{\bm{\mu}} \\
  {{{\bm{\mu}}^{\text{T}}}}&{\mathbf{1}}
\end{array}} \right],
\end{align}
where ${\bm{\mu }} \in {\mathbb{R}^n}$ and ${\mathbf{\Sigma }} \in {\mathbb{S}^n}$  are the mean vector  and covariance matrix  of  random vector ${\bm{\xi }}$, respectively.

{Let define the continuous quadratic function $ L\left( {\Delta {\bf{h}}} \right) = \Delta\mathbf{h}^\mathrm{T}(-\mathbf{V})\Delta\mathbf{h}+2\Delta\mathbf{h}^\mathrm{T}(-\mathbf{V})\hat{\mathbf{h}}+\hat{\mathbf{h}}^\mathrm{T}(-\mathbf{V})\hat{\mathbf{h}}+\frac{\delta}{P_{\mathrm{c}}}$. By Lemma $2$, the worst-case chance constraint in \eqref{CRaV1} can be computed by the optimization problem as similarly as the problem \eqref{eqn:lemma2}. Then, according to the Lemma $1$, the problem can be equivalent to the following CVaR constraints:}
\begin{subequations}
\begin{align}
 &\beta  + \frac{1}{{{P_{{\rm{out}}}}}}{\rm{Tr}}\left( {{\bf{\Omega M}}} \right) \le 0,\label{CRaV_constraints1}\\
  &{\bf{M}} - \left[ {\begin{array}{*{20}{c}}
-{\bf{V}}&-{{{\bf{V}}^{\text{T}}}{\bf{\hat h}}}\\
-{{{{\bf{\hat h}}}^{\text{T}}}{\bf{V}}}&-{{{{\bf{\hat h}}}^{\text{T}}}{\bf{V\hat h}} + \frac{\delta }{{{P_{\rm{c}}}}} - \beta }
\end{array}} \right]\succeq {\bf{0}},\label{CRaV_constraints2}\\
&{\bf{M}}\succeq {\bf{0}},{\bf{M}} \in {\mathbb{S}}{^4}, \label{CRaV_constraints3}
\end{align}
\end{subequations}
where ${\bf{M}}$ and $\beta$ are two auxiliary variables, and
\[{\bf{\Omega }} = \left[ {\begin{array}{*{20}{c}}
{{\bf{D}} + {\bm{\mu }}{{\bm{\mu }}^{\text{T}}}}&{\bm{\mu }}\\
{\bm{\mu }}^{\text{T}}&{\bf{1}}
\end{array}} \right].\]
Therefore,  the original distributionally chance-constrained problem \eqref{tradeoff_originalPro1}  can be reformulated as follows
\begin{subequations}\label{BDC}
\begin{align}
 \mathop {\min }\limits_{{P_{\rm{p}}},{P_{\rm{c}}},{\bf{V}},{\bf{M}},\beta } \;~&{\rm{Tr}}\left( {{\bf{J}}_{\rm{p}}^{ - 1}} \right)\label{BDC_Obj}\\
{\rm{s.t.}} ~&\eqref{robust_cons_Ptotal},\eqref{constraint_Pp},
\eqref{constraint_Pc},\eqref{semidefinite V2},\eqref{CRaV_constraints1},\eqref{CRaV_constraints2},
\eqref{CRaV_constraints3}.\nonumber
\end{align}
\end{subequations}

{Note that, problem \eqref{BDC} is still non-convex  given that the    optimization variables $P_{\rm{p}}$ and
${{\bf{V}}}$ are coupled together   in constraint \eqref{CRaV_constraints1}}.
{However, the problem  \eqref{BDC} can be decomposed  into two convex subproblems with two decoupling variables blocks: $\left\{ {{P_{\rm{c}}},{\bf{V}},{\bf{M}},\beta } \right\}$ and ${P_{\rm{p}}}$, respectively. It means that when one of blocks is fixed, the problem becomes convex in the remaining block of variables, which is called the multi-convex problem\cite{Xu2013_SIAM}. To solve this kind of multi-convex problem,} we  propose an efficient BCD algorithm  \cite{1999Nonlinear} for robust VLPC design with variables coupling{, which can  guarantee to globally converge to the stationary point \cite{Grippof1999,Grippo2000}.}
Then,   at every iteration,
the two convex subproblems, i.e.,  VLP subproblems and  VLC subproblems,   are alternatively optimized with respect to one block
variable  while the remaining blocks are held fixed.
 More specifically, for the $k$th iteration,    the VLP and VLC   subproblems  are optimized   follows.

\subsubsection{\textbf{VLP subproblem}} For fixing variables $P_\mathrm{c}^{\left( {k-1} \right)}$, the positioning power
  $P_{\rm{p}}^{\left( {k} \right)}$   is updated  via solving  the following  convex VLP subproblem
  {
\begin{align}\label{sub_problem2}
\mathop {\min }\limits_{{P_{\rm{p}}}}&~ {\rm{Tr}}\left( {{\bf{J}}_{\rm{p}}^{ - 1}} \right)\\
{\rm{s}}.{\rm{t}}.
&~
\eqref{robust_cons_Ptotal},~\eqref{constraint_Pp},\nonumber
\end{align}
}
 which can be solved using the interior point methods, such as CVX \cite{cvx}.

\subsubsection{\textbf{VLC subproblem}}
 With given positioning power $P_{\rm{p}}^{\left( {k} \right)}$,
the block variables
 $\left\{ P_\mathrm{c}^{\left( {k} \right)},{\bf{V}}^{\left( {k }\right)},\mathbf{M}^{\left(k\right)},\beta^{\left(k\right)} \right\}$ are   updated by solving  the following  convex VLC subproblem
\begin{align}\label{sub_problem1}
\mathop {\min }\limits_{{P_{\rm{c}}},{\bf{V}},{\bf{M}},\beta } &~{P_{\rm{c}}}\\
{\rm{s}}.{\rm{t}}.&~
\eqref{constraint_Pc}, ~\eqref{semidefinite V2},
~\eqref{CRaV_constraints1},~\eqref{CRaV_constraints2},
~\eqref{CRaV_constraints3}.\nonumber
\end{align}

 In summary, the overall BCD   algorithm for robust VLPC design  is listed in
Algorithm 1.
The solution of the  BCD Algorithm 1 is a
stationary point of the joint optimization problem   \eqref{tradeoff_originalPro1} \cite{Hu_TSP_2018,Wang_TSP_2015}.
Note that, due to the SDR, the rank of ${{{\bf{V}}^{{\left( k \right)}}}}$ may not be $1$.
For ${\rm{rank}}\left( {{{\bf{V}}^{{\left( k \right)}}}} \right)=1$, the optimal beamformer
$\bf{v}$ can be calculated  by eigenvalue decomposition.
 When
${\rm{rank}}\left( {{{\bf{V}}^{{\left( k \right)}}}} \right)>1$,
 we can calculate a high-quality feasible solution $\bf{v}$ of problem \eqref{sub_problem1} based on the Gaussian randomization procedure \cite{Semidefinite2010Luo}. { Meanwhile, the two SDP problem can be efficiently solved with a worst case complexity $\mathcal{O}\left(\max\left\{m,n\right\}^{4}n^{0.5}\log\delta^{-1}\right)$, where $n$ is the problem size $n$, $m$ denotes the number of constraints $m$, and $\delta$ represents the accuracy of SDP \cite{Semidefinite2010Luo}. And the proposed BCD algorithm has a sub-linear convergence rate, $\mathcal{O}\left(\frac{1}{k}\right)$, where $k$ is the index of iteration \cite{Cartis2010}.}

\begin{algorithm}[htbp]
    \caption{Block Coordinate Descent Algorithm for Robust VLPC Design}
    \begin{algorithmic}[1]
        \Require Initialize ${\bar r}$, ${P_{{\rm{out}}}}$, $P_{\text{p}}^{\left( 0 \right)}$, $k=0$ and
        set the tolerance of accuracy  $\epsilon >0 $;
        \Repeat
        \State $k \leftarrow k + 1$;
        \State Update $\left\{ P_c^{\left( {k} \right)}, {{\bf{V}}^{\left( {k } \right)}},\mathbf{M}^{\left(k\right)},\beta^{\left(k\right)} \right\}$ by solving  VLC subproblem \eqref{sub_problem1} with fixed    $P_{\rm{p}}^{\left( {k - 1} \right)}$;
        \State Update $P_{\rm{p}}^{\left( {k} \right)}$   by    solving VLP subproblem \eqref{sub_problem2}  with given $\left\{ P_c^{\left( {k} \right)},{{\bf{V}}^{\left( {k } \right)}},\mathbf{M}^{\left(k\right)},\beta^{\left(k\right)} \right\}$;
        \Until{ $\left| {{\rm{Tr}}{{\left( {{\bf{J}}_{\rm{p}}^{ - 1}} \right)}^{\left( k \right)}} - {\rm{Tr}}{{\left( {{\bf{J}}_{\rm{p}}^{ - 1}} \right)}^{\left( {k - 1} \right)}}} \right| \le  \epsilon$;}
        \Ensure  $P_{\rm{p}}^{\left( {k} \right)}$, $P_{\rm{c}}^{\left( {k} \right)}$ and ${{{\bf{V}}^{{\left( k \right)}}}}$.
    \end{algorithmic}
\end{algorithm}

\section{ Simulation results}\label{section_v}

In this section,  we present simulation results to  evaluate the effectiveness of the proposed VLPC system  design.
Consider a  VLPC system in a room with size $\left( {5 \times 5 \times 3  {m^3}} \right)$, where one corner of the room is the origin $\left( {0,0,0} \right)$ of the Cartesian coordinate system $\left( {X,Y,Z} \right)$.
Assume that the LED location  is  $\left( {2.5,2.5,3} \right)$ and the MU is equipped with $M=3$ PDs.

\begin{figure}[htbp]
    \centering	\includegraphics[width=0.3\textwidth]{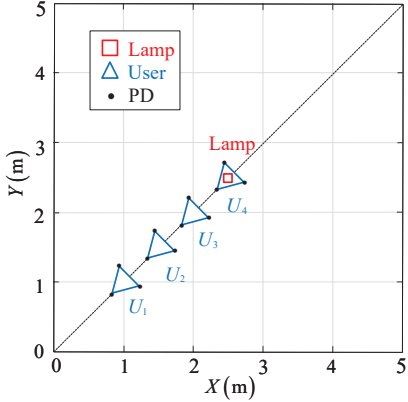}
    \caption{Locations  of MU and LED.}
    \label{PD_location_diagram}
\end{figure}

Moreover,  as shown in Fig. \ref{PD_location_diagram},  we  verify the performance of the proposed optimization method  for four different horizontal locations of  the MU, i.e.,
${U_1}\left( {1,1,z_u} \right)$,
${U_2}\left( {1.5,1.5,z_u} \right)$,
${U_3}\left( {2,2,z_u} \right)$ and
${U_4}\left( {2.5,2.5,z_u} \right)$,
where the PDs  are arranged according to an equilateral triangle with side length $L$.
The  other simulation parameters are summarized in Table \ref{Tab:positioning_method}.

\begin{table}[htpb]
    \centering
    \caption{Basic Simulation Parameters}
    \begin{tabular}{|l|l|}
        \hline
        Parameters & Value   \\
        \hline
        FoV, ~${\psi _{{\rm{FoV}}}}$ & ${90^ \circ }$   \\
        \hline
        Detector area of PD,~ $A_{\rm{PD}}$ & $1{\rm{c}}{{\rm{m}}^2}$   \\
        \hline
        Half power angle,~${\theta _{1/2}}$&${60^ \circ }$ \\
        \hline
        Gain of an optical filter,~$T_f $ & 1\\
        \hline
        Gain of an optical concentrator,~$g $ & 1\\
        \hline
        DC bias,~$I_{\rm{DC}} $ & 2\\
        \hline
        Peak amplitude,~$A$ & 0.007\\
        \hline
        Maximum optical power,~$P_{\mathrm{o}}^{\max }$ & 8W\\
        \hline
        Maximum electrical  power,~$P_{\mathrm{e}}^{\max }$ & 16W\\
        \hline
        Bandwidth,~$B$ & $20$MHz\\
        \hline
        Noise PSD of positioning signal,~$\sigma _{\rm{p}}^2$
        & ${{{{10}^{ - 21}}{\rm{Watts}}} \mathord{\left/
                {\vphantom {{{{10}^{ - 21}}{\rm{Watts}}} {{\rm{Hz}}}}} \right.
                \kern-\nulldelimiterspace} {{\rm{Hz}}}}$ \\
        \hline
        Noise PSD of data signal,~$\sigma _{\rm{c}}^2$
        & ${{{{10}^{ - 21}}{\rm{Watts}}} \mathord{\left/
                {\vphantom {{{{10}^{ - 21}}{\rm{Watts}}} {{\rm{Hz}}}}} \right.
                \kern-\nulldelimiterspace} {{\rm{Hz}}}}$ \\
        \hline
    \end{tabular}
    \label{Tab:positioning_method}
\end{table}

\subsection{Positioning Performance }

First of all, it should be noted that the positioning error in this section is the root mean square error (RMSE),  which is the error between the average estimated value obtained from multiple measurements and the true value. The positioning symbol with normalized power is considered, i.e., $\varepsilon=1$.
The received SNR is defined as the received SNR at the $1$th PD, i.e,  ${\rm{SNR = 10lg}}\frac{{{P_{\rm{p}}}h_1^2}}{{B\sigma _{\rm{p}}^2}}$.

\begin{figure}[htbp]
      \centering	\includegraphics[width=0.4\textwidth]{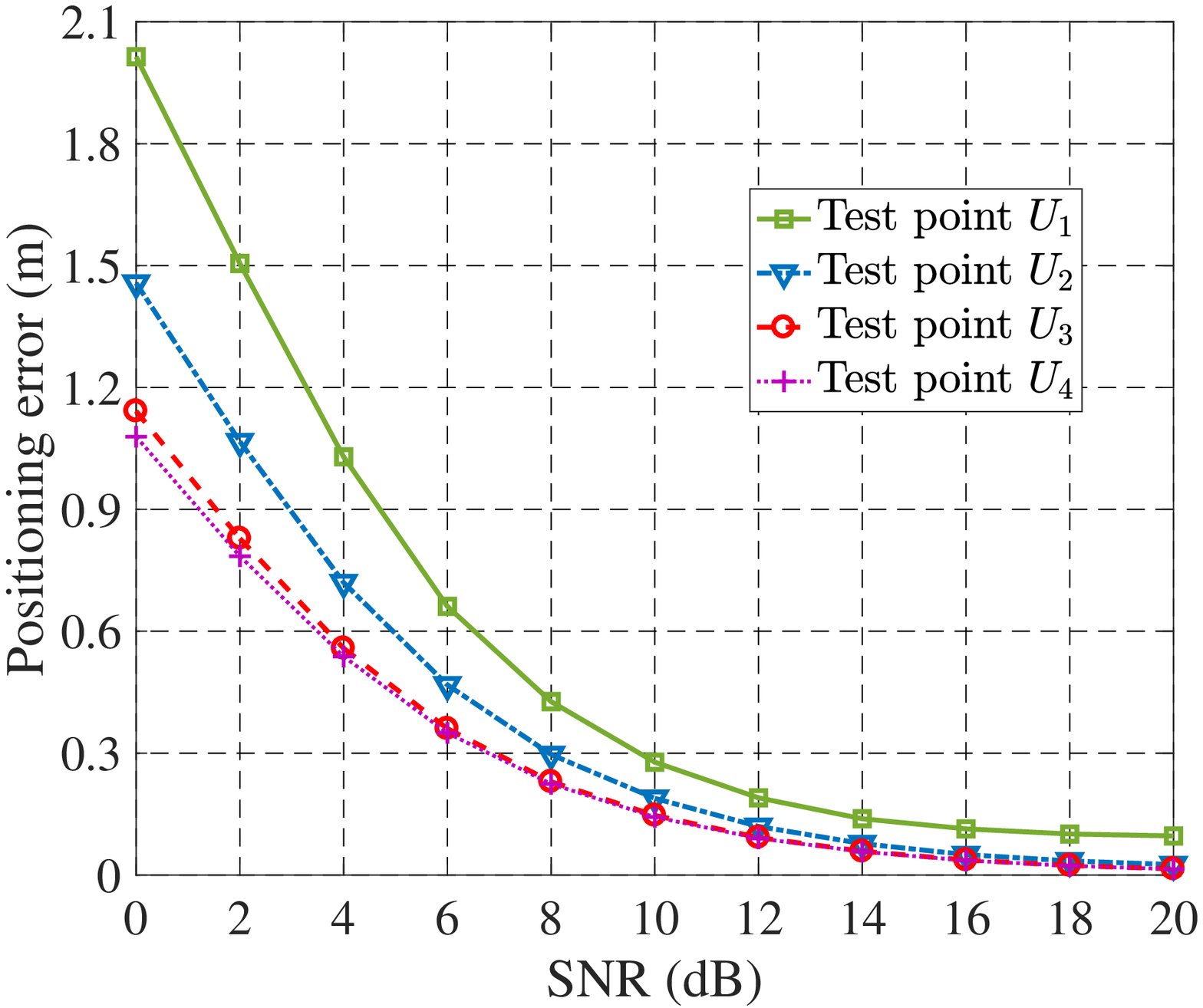}
      \vskip-0.2cm\centering {\footnotesize (a)}

       \centering \includegraphics[width=0.4\textwidth]{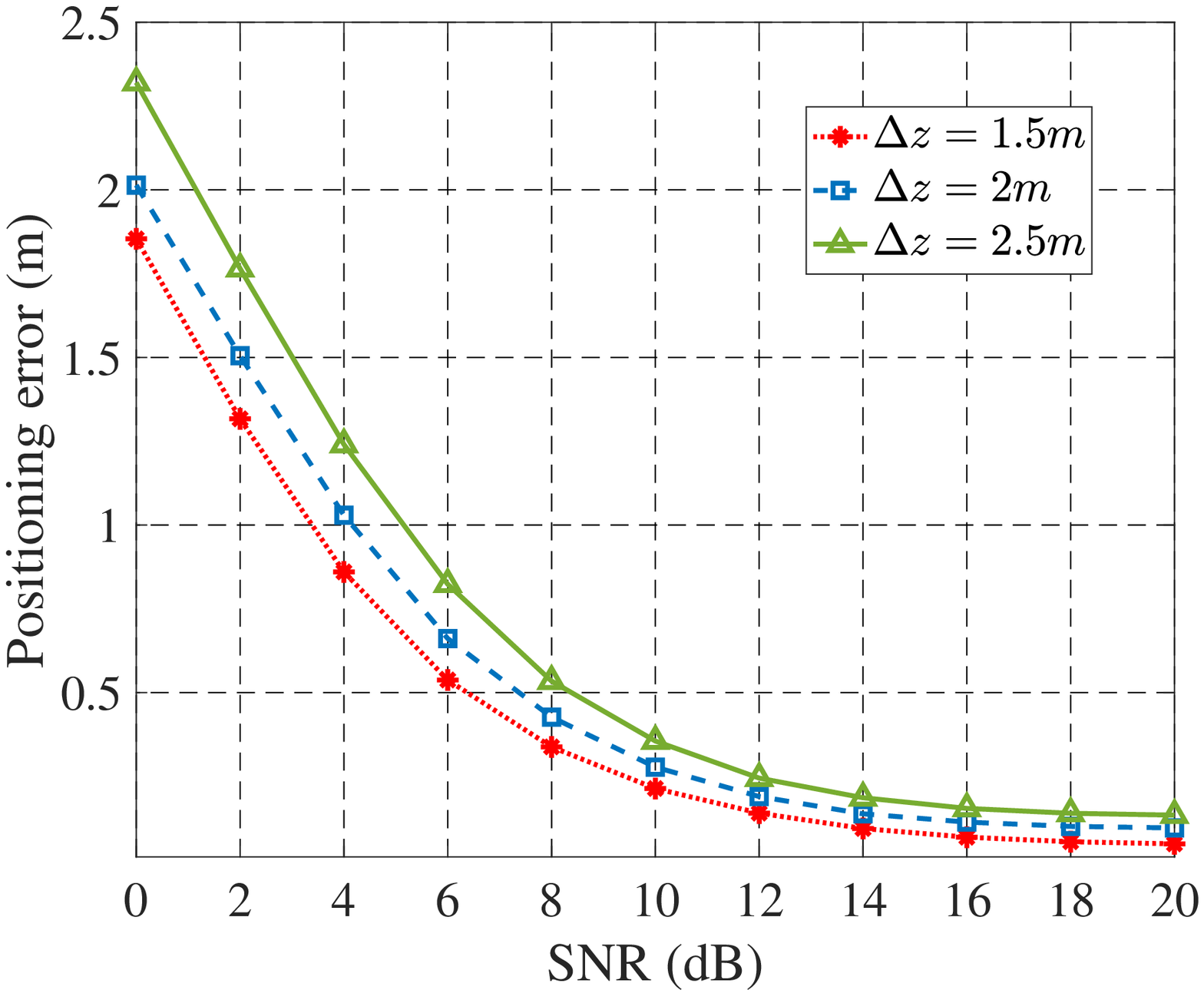}
     \vskip-0.2cm\centering {\footnotesize (b)}

    \caption{
(a) Positioning error versus SNR when the test point is chosen  at different locations, where $L=0.1m$;
(b) Positioning error versus SNR when the transceiver height difference $\Delta z={z_l} - {z_u}$ is different,  where  $L=0.1m$.}
  \label{L10cm_compare_point&h}
\end{figure}
Fig. \ref{L10cm_compare_point&h} (a) illustrates the positioning error at the four test points versus received SNR,
where $z_u=1m$.
We may observe that the positioning error decreases rapidly at first and then slowly, and finally converge to  a constant as  SNR  increases.
This is because as SNR  increases, the influence of the noise decreases.
For high SNR, the localization performance is negligibly affected by the noise, but still affected by the NLOS link.
In addition,  it can be seen from the Fig. \ref{L10cm_compare_point&h} (a) that
the positioning error at test points $U_1$, $U_2$, $U_3$ and $U_4$ gradually decreases at the same SNR.
This is because  $U_4$ is the closest to the lamp, while $U_1$ is the farthest.
Therefore,  when the total transmit power   is constant, the SNR at $U_4$, $U_3$, $U_2$ and $U_1$ decreases.

In Fig. \ref{L10cm_compare_point&h} (b), we plot the  positioning error at   test point $U_1$  versus SNR,
for different MU heights.
The shapes of the curves are similar to Fig. 4 (a), and the reasons are also similar.

\begin{figure}[htbp]
      \centering	\includegraphics[width=0.4\textwidth]{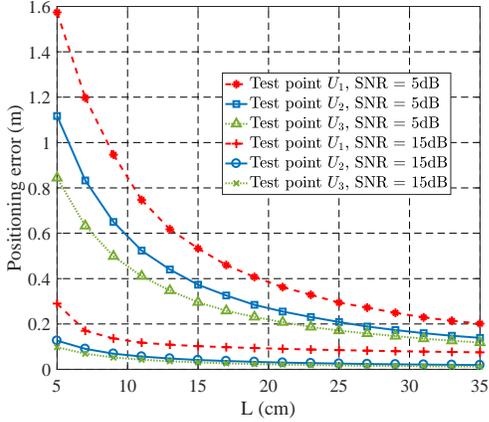}
    \caption{Positioning error versus $L$(cm), where $\Delta z=2m$.}
  \label{snr15and5_compare_L}
\end{figure}
Fig. \ref{snr15and5_compare_L} shows the positioning error  versus side length $L$ at high and low SNR.
As we can   see, with the increase of relative distance $L$, the positioning error first decreases rapidly and then slowly when the SNR=5dB, while the position error first decreases slowly and then remains unchanged when the SNR=15dB.
This is because
the larger the  relative distance $L$ is, the greater the difference in  signal intensity received by each PD  will be.
Thus, the solution of the nonlinear equations in \eqref{fangcheng_2} will be more accurate,
especially at low SNR, and  the impact of the  signal strength difference on accuracy is more obvious.
At the same time, when the  relative distance $L$ reaches a certain value, additional increases will not help.
In addition, when $L$ is constant, the positioning error at $U_1$, $U_2$ and $U_3$ still decreases sequentially,
and the gap between the  positioning errors at $U_2$ and $U_3$ becomes smaller with increasing $L$.

 \subsection{Communication  Performance}

To  evaluate the communication performance, we first introduce the non-robust VLPC design scheme,  which  ignores the CSI uncertainty  $\Delta {\bf{h}}$,
and the estimated CSI $\widehat {\bf{h}}$ is viewed as the perfect CSI ${\bf{h}}$.
Moreover, in our simulations,  we choose the following basic parameters:
length of positioning subframe $T_{\mathrm{p}}=0.12\,\mathrm{sec}$
side length $L = 0.1\,\mathrm{m}$ and the test point  $U_3$  with $z_u=1.5\,\mathrm{m}$.

\begin{figure}[htbp]
    \centering
\begin{minipage}[b]{0.4\textwidth}
      \centering
\includegraphics[width=\textwidth]{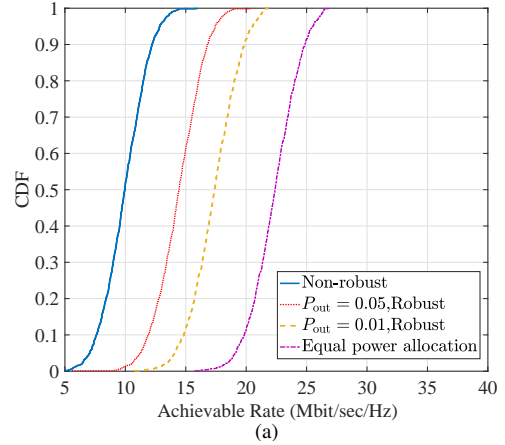}
      \vskip-0.2cm\centering {\footnotesize (a)}
    \end{minipage}\hfill
    \begin{minipage}[b]{0.4\textwidth}
      \centering
\includegraphics[width=\textwidth]{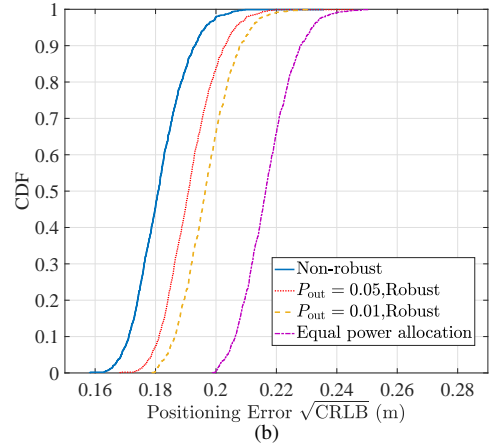}
      \vskip-0.2cm\centering {\footnotesize (b)}
    \end{minipage}\hfill
  \caption{In the non-robust VLPC design, the robust VLPC design with outage probabilities ${P_{{\rm{out}}}} = 1\% $ and $5\%$, and the equal power allocation design: (a)~ CDF of achievable  rate; (b)~ CDF of CRLB.}
  \label{CDF of rate}
\end{figure}
Fig. \ref{CDF of rate} (a) depicts the CDF of achievable  rate  of the non-robust VLPC design, the robust VLPC design with the maximum tolerated outage probabilities ${P_{{\rm{out}}}} = 1\% $ and $5\%$, and the equal power allocation design (${P_{\rm{p}}} = {P_{\rm{c}}}$),
 where  the total power is ${P_{{\rm{T}}}} = 10W$.
It can be seen that the  outage probability   of   the  non-robust VLPC design is $50\% $, which  significantly
exceeds the maximum tolerated outage probability requirement.  On the other hand, the outage probability of  the proposed  robust VLPC design   is lower than $5\% $, which  meets the outage probability requirement.
Fig. \ref{CDF of rate} (b) depicts the CDF of CRLB  with the same  parameters as Fig. \ref{CDF of rate} (a).
As can be seen from Fig. \ref{CDF of rate} (b),
the positioning error of the robust VLPC design with  ${P_{{\rm{out}}}} = 1\% $  is lager than that of the design with ${P_{{\rm{out}}}} = 5\% $.
Combined with Fig. \ref{CDF of rate} (a), the robust VLPC design allocates more power to communication than non-robust design under  the premise of minimizing positioning accuracy in order to meet the minimum rate requirements of the system.
Therefore, when the total power is limited, the positioning power decreases correspondingly, resulting in the increase of positioning error.
Compared with the equal power allocation scheme, the robust VLPC scheme allocates less power to communication under the condition of satisfying the rate requirement, resulting in less positioning error.
Thus, Fig. \ref{CDF of rate}    demonstrates the effectiveness of our proposed robust VLPC design.

\begin{figure}[htbp]
  \centering
 \subfigure[]{
\includegraphics[width=0.4\textwidth]{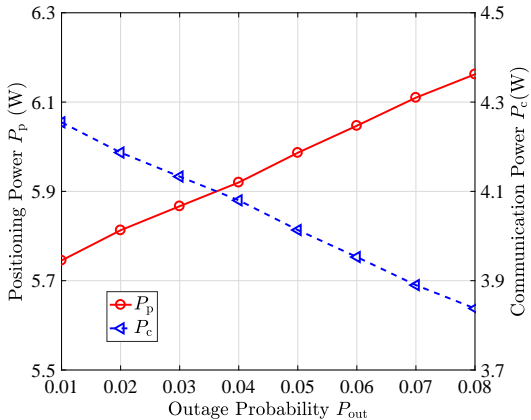}
   }
    \subfigure[]{
\includegraphics[width=0.4\textwidth]{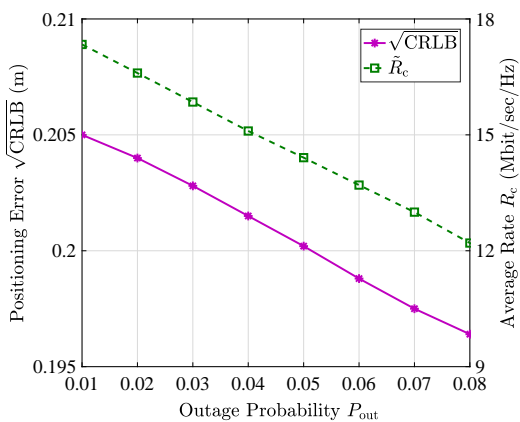}
  }
   \centering
  \caption{(a)~ Power allocation of robust VLPC design  versus outage probability ${P_{{\rm{out}}}}$; (b)~ CRLB and average communication rate ${{\tilde R}_{\rm{c}}}$ of robust VLPC design  versus outage probability ${P_{{\rm{out}}}}$.}
  \label{P_allocation_Pout}
\end{figure}

Fig. \ref{P_allocation_Pout} (a) shows the   positioning power ${P_{\rm{p}}}$  and the communication power ${P_{\rm{c}}}$  of the robust VLPC design versus the  outage probability ${P_{{\rm{out}}}}$  with  $\bar r = 10 {\rm{Mbit}}/\mathrm{sec} $ and ${P_{{\rm{T}}}} = 10W$.
From Fig. \ref{P_allocation_Pout} (a),
with increasing outage probability ${P_{{\rm{out}}}}$,
the positioning power ${P_{\rm{p}}}$ increases, while  the communication power ${P_{\rm{c}}}$ decreases.
This is because  as the  outage probability ${P_{{\rm{out}}}}$ decreases,   the probability that the communication rate is the threshold $\bar r $ decreases,  and
the robust design becomes  more conservative.
Moreover, under the same     simulation conditions as Fig. \ref{P_allocation_Pout} (a),   Fig. \ref{P_allocation_Pout} (b) depicts   the  CRLB    and average communication rate ${{\tilde R}_{\rm{c}}}$ of  the robust VLPC design versus the outage probability ${P_{{\rm{out}}}}$.
We observe that,
as the outage probability ${P_{{\rm{out}}}}$ increases,
the CRLB decreases, and the average communication rate ${{\tilde R}_{\rm{c}}}$  also decreases.
This is because  for a given total power, the communication power ${P_{{\rm{c}}}}$  and the positioning power ${P_{{\rm{p}}}}$ are both related to the communication rate, and there exists a tradeoff between them.

\begin{figure}[htbp]
  \centering
 \subfigure[]{
\includegraphics[width=0.4\textwidth]{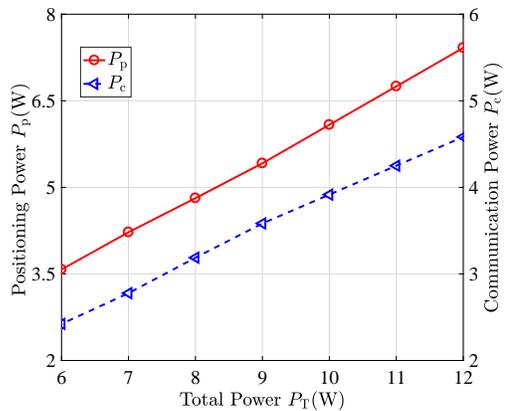}
   }
    \subfigure[]{
\includegraphics[width=0.4\textwidth]{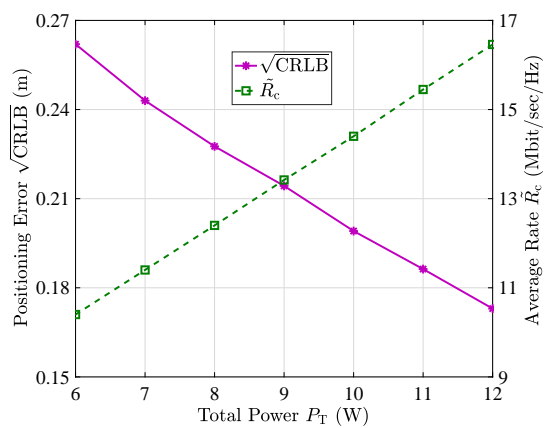}
  }
   \centering
  \caption{(a)~ Power allocation of robust VLPC design versus total power ${P_{{\rm{T}}}}$; (b)~ CRLB and average   rate ${{\tilde R}_{\rm{c}}}$ of robust VLPC design versus total power ${P_{{\rm{T}}}}$.}
  \label{P_allocation_Pt}
\end{figure}

Fig. \ref{P_allocation_Pt} show the influence  of the total power ${P_{{\rm{T}}}}$ on the robust VLPC design.
Fig. \ref{P_allocation_Pt} (a) shows
the optimized power allocation
versus ${P_{{\rm{T}}}}$ with  ${P_{{\rm{out}}}} = 5\%$ and  $\bar r = 5 {\rm{Mbit}}/\mathrm{sec}$.
From Fig. \ref{P_allocation_Pt} (a), we can  observe that as the total power ${P_{{\rm{T}}}}$ increases, both the positioning power ${P_{\rm{p}}}$  and the communication power ${P_{\rm{c}}}$
increase because more power can be allocated for both positioning and communication power to meet the positioning performance requirements and rate constraints.  In  addition,
Fig. \ref{P_allocation_Pt} (b) shows the  CRLB and average   rate ${{\tilde R}_{\rm{c}}}$
versus
${P_{{\rm{T}}}}$ with the same   parameters as Fig. \ref{P_allocation_Pt} (a).
From Fig. \ref{P_allocation_Pt} (b),
we can observe that as the  total power ${P_{{\rm{T}}}}$ increases, the CRLB decreases and the average communication rate ${{\tilde R}_{\rm{c}}}$  increases.
This is intuitive since higher total available power improves both the positioning and communication.

\begin{figure}[htbp]
    \centering
    \subfigure[]{        \includegraphics[width=0.4\textwidth]{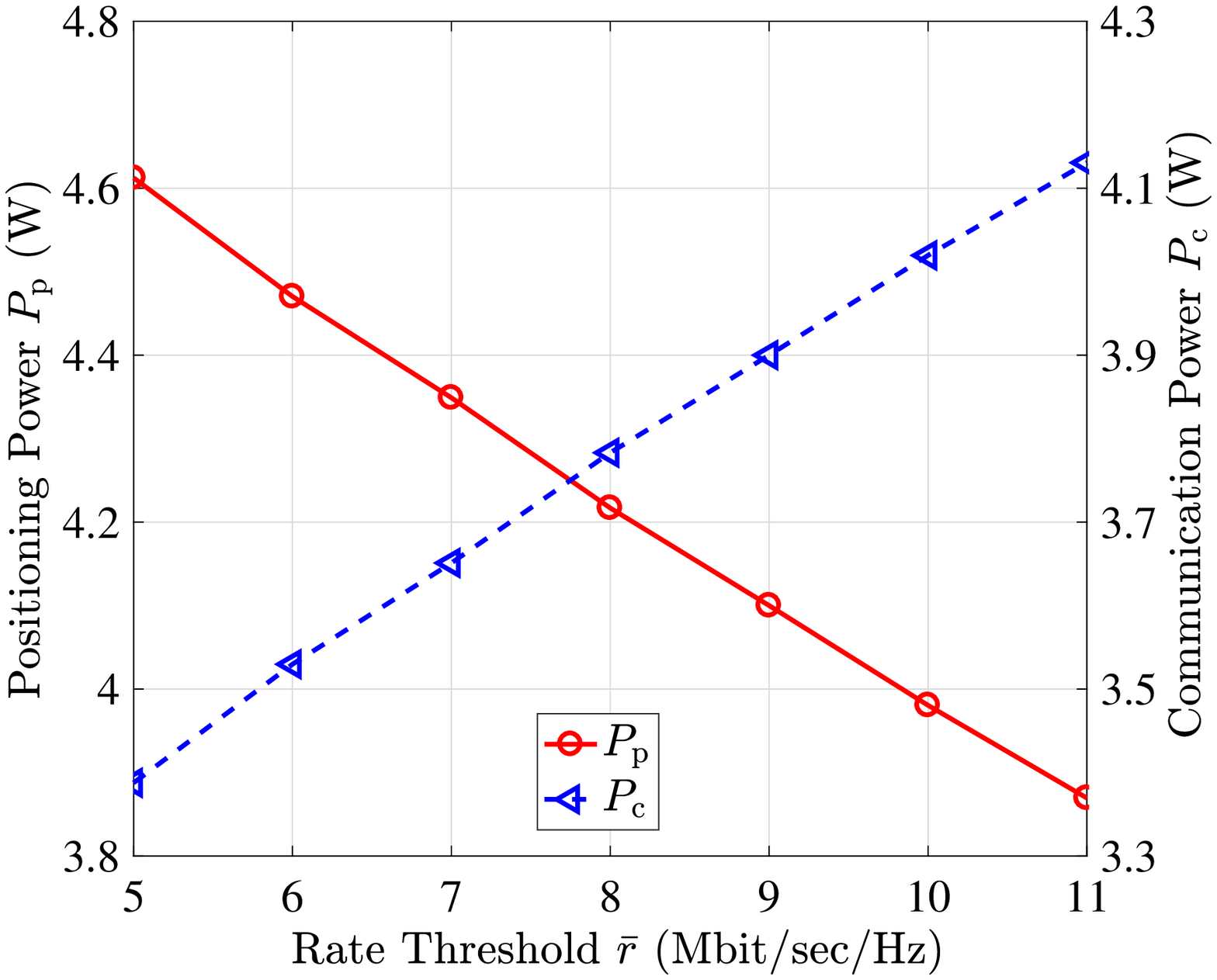}    }
    \subfigure[]{        \includegraphics[width=0.4\textwidth]{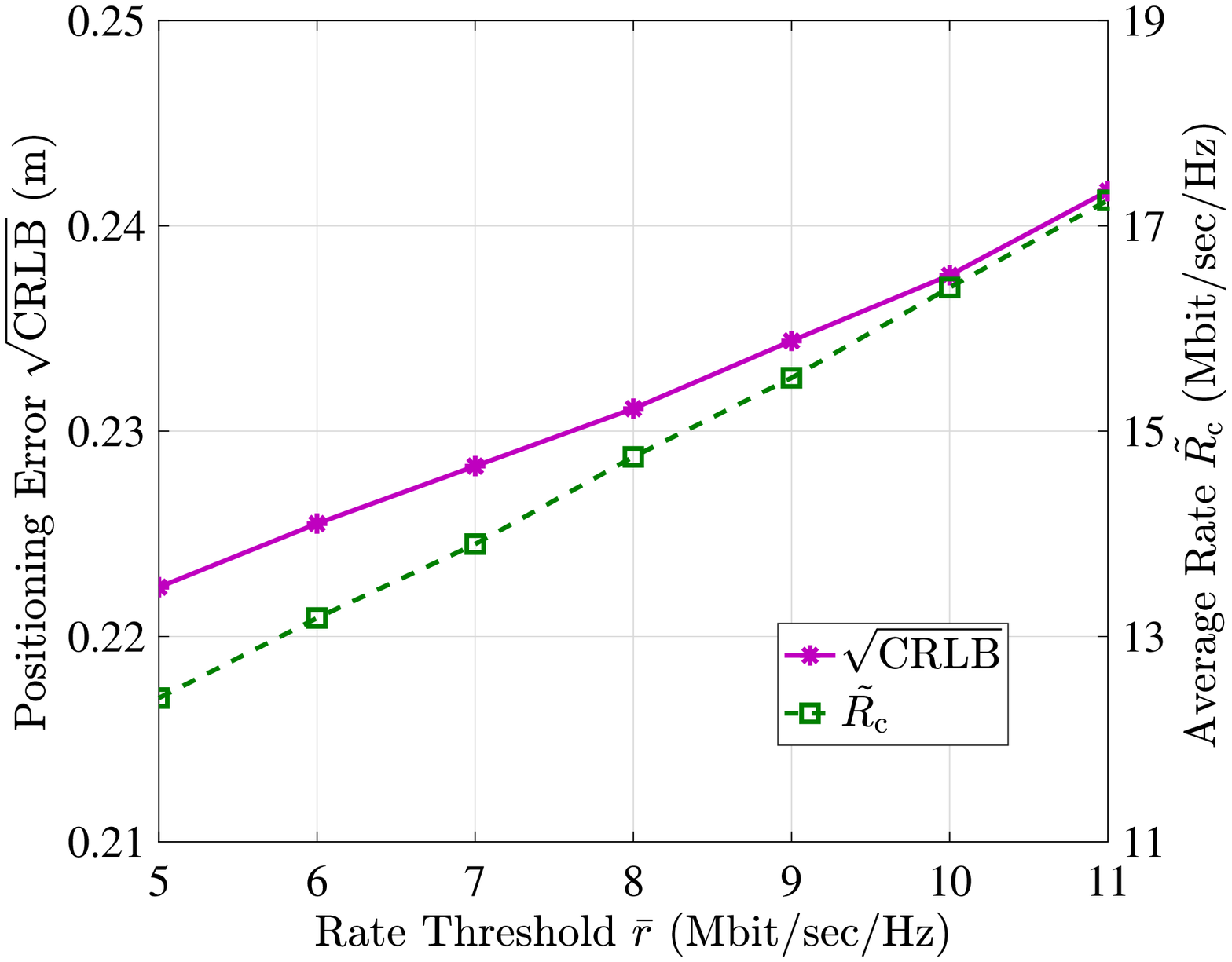}    }
    \centering
    \caption{(a)~ Power allocation of robust VLPC design versus rate threshold $\bar r$; (b)~ CRLB and average   rate ${{\tilde R}_{\rm{c}}}$ of robust VLPC design versus rate threshold $\bar r$.}
    \label{P_allocation_rbar}
\end{figure}

Fig. \ref{P_allocation_rbar} show the influence of the rate thresholds $\bar r$ on the robust VLPC design,
where ${P_{{\rm{T}}}} = 8W$ and ${P_{{\rm{out}}}} = 5\% $.
Fig. \ref{P_allocation_rbar} (a) shows the optimized  power allocation versus different rate thresholds $\bar r$.
We can see that the allocated positioning power ${P_{\rm{p}}}$ decreases  while ${P_{\rm{c}}}$ increases as the rate threshold $\bar r$ increases.
This is because  the robust VLPC system  needs more communication power to meet the  rate threshold  $\bar r$.
Moreover, Fig. \ref{P_allocation_rbar} (b) shows the
CRLB and average   rate ${{\tilde R}_{\rm{c}}}$   versus  $\bar r$.
It  can be seen that  as the rate threshold $\bar r$ increases, the CRLB increases because the positioning power ${P_{\rm{p}}}$ decreases as  $\bar r$ increases.
In addition, the average rate  increases because the communication power ${P_{\rm{c}}}$  increases as  $\bar r$ increases.

\section{Conclusion}\label{section_vi}

 In this paper, we
 reveal the intrinsic relationship between VLP and VLC based on the relationship between   CSI and location, i.e., the positioning information can be used to estimate the CSI.
Then, both the CRLB  for  VLP  and the achievable rate of VLC   are  derived.
Furthermore, a  robust    power allocation scheme  is proposed under   practical optical constraints, and
QoS requirements.
To tackle  the rate outage constraints, the worst-case distribution of  the  CVaR  is conservatively   approximated to a more tractable form.
Then, we propose a BCD  Algorithm for     robust VLPC design, in which the VLP and VLC sub-problems  are iteratively optimized.
Finally, our simulation results demonstrate the effectiveness of the proposed VLPC scheme for both localization and communications.

\begin{appendices}

\section{Derivation of Equation \eqref{FIM}}\label{tuidao_FIM}

The Fisher Information matrix (FIM)  is given by
\begin{align}
{\mathbf{J}}_{\rm{p}} = \left[ {\begin{array}{*{20}{c}}
  { - \mathbb{E}\left\{ {\frac{{{\partial ^2}\Lambda \left( {\mathbf{u}} \right)}}{{\partial x_u^2}}} \right\}}&{ - \mathbb{E}\left\{ {\frac{{{\partial ^2}\Lambda \left( {\mathbf{u}} \right)}}{{\partial {x_u}\partial {y_u}}}} \right\}}&{ - \mathbb{E}\left\{ {\frac{{{\partial ^2}\Lambda \left( {\mathbf{u}} \right)}}{{\partial {x_u}\partial {z_u}}}} \right\}} \\
  { - \mathbb{E}\left\{ {\frac{{{\partial ^2}\Lambda \left( {\mathbf{u}} \right)}}{{\partial {y_u}\partial {x_u}}}} \right\}}&{ - \mathbb{E}\left\{ {\frac{{{\partial ^2}\Lambda \left( {\mathbf{u}} \right)}}{{\partial y_u^2}}} \right\}}&{ - \mathbb{E}\left\{ {\frac{{{\partial ^2}\Lambda \left( {\mathbf{u}} \right)}}{{\partial {y_u}\partial {z_u}}}} \right\}} \\
  { - \mathbb{E}\left\{ {\frac{{{\partial ^2}\Lambda \left( {\mathbf{u}} \right)}}{{\partial {z_u}\partial {x_u}}}} \right\}}&{ - \mathbb{E}\left\{ {\frac{{{\partial ^2}\Lambda \left( {\mathbf{u}} \right)}}{{\partial {z_u}\partial {y_u}}}} \right\}}&{ - \mathbb{E}\left\{ {\frac{{{\partial ^2}\Lambda \left( {\mathbf{u}} \right)}}{{\partial z_u^2}}} \right\}}
\end{array}} \right].
\end{align}

Then, based on the first  partial derivatives of the likelihood function in \eqref{log_likehood}, we have
\begin{subequations}
\begin{align}
&\frac{{\partial \Lambda \left( {{{\mathbf{u}}}} \right)}}{{\partial {x_u}}} =  - \frac{1}{{{\sigma _{\rm{p}}^2}}}\sum\limits_{i = 1}^M {\int_0^{T_{\rm{p}}} {\left( {{h_i}x_{\text{p}}^2\left( t \right) - {y_{{\text{p}},i}}\left( t \right){x_{\text{p}}}\left( t \right)} \right)\frac{{\partial {h_i}}}{{\partial {x_u}}}\,\mathrm{d}t} }  ,\\
&\frac{{\partial \Lambda \left( {{{\mathbf{u}}}} \right)}}{{\partial {y_u}}} =  - \frac{1}{{{\sigma _{\rm{p}}^2}}}\sum\limits_{i = 1}^M {\int_0^{T_{\rm{p}}} {\left( {{h_i}x_{\text{p}}^2\left( t \right) - {y_{{\text{p}},i}}\left( t \right){x_{\text{p}}}\left( t \right)} \right)\frac{{\partial {h_i}}}{{\partial {y_u}}}\,\mathrm{d}t} },\\
&\frac{{\partial \Lambda \left( {{{\mathbf{u}}}} \right)}}{{\partial {z_u}}} =  - \frac{1}{{{\sigma ^2}}}\sum\limits_{i = 1}^M {\int_0^{T_{\rm{p}}} {\left( {{h_i}x_{\text{p}}^2\left( t \right) - {y_{{\text{p}},i}}\left( t \right){x_{\text{p}}}\left( t \right)} \right)\frac{{\partial {h_i}}}{{\partial {z_u}}}\,\mathrm{d}t} }.
\end{align}
\end{subequations}

Furthermore,    according to the  second partial derivatives of the likelihood function,   we   obtain
\begin{subequations}\label{partial2}
\begin{align}
\frac{{{\partial ^2}\Lambda \left( {{{\mathbf{u}}}} \right)}}{{\partial x_u^2}}& =  - \frac{1}{{\sigma _{\rm{p}}^2}}\sum\limits_{i = 1}^M {\int_0^{{T_{\rm{p}}}} {\left( {\frac{{\partial {h_i}}}{{\partial {x_u}}}\frac{{\partial {h_i}}}{{\partial {x_u}}}x_{\rm{p}}^2\left( t \right) + {h_i}\frac{{{\partial ^2}{h_i}}}{{\partial x_u^2}}x_{\rm{p}}^2\left( t \right)} \right.} }\nonumber\\
&\quad\quad\quad\quad\quad\quad\quad\left. { - {y_{{\rm{p}},i}}\left( t \right){x_{\rm{p}}}\left( t \right)\frac{{{\partial ^2}{h_i}}}{{\partial x_u^2}}} \right)\,\mathrm{d}t,\\
\frac{{{\partial ^2}\Lambda \left( {{{\mathbf{u}}}} \right)}}{{\partial y_u^2}} &=   - \frac{1}{{\sigma _{\rm{p}}^2}}\sum\limits_{i = 1}^M {\int_0^{{T_{\rm{p}}}} {\left( {\frac{{\partial {h_i}}}{{\partial {y_u}}}\frac{{\partial {h_i}}}{{\partial {y_u}}}x_{\rm{p}}^2\left( t \right) + {h_i}\frac{{{\partial ^2}{h_i}}}{{\partial y_u^2}}x_{\rm{p}}^2\left( t \right)} \right.} } \nonumber\\
&\quad\quad\quad\quad\quad\quad\quad\left. { - {y_{{\rm{p}},i}}\left( t \right){x_{\rm{p}}}\left( t \right)\frac{{{\partial ^2}{h_i}}}{{\partial y_u^2}}} \right)\,\mathrm{d}t,\\
\frac{{{\partial ^2}\Lambda \left( {\bf{u}} \right)}}{{\partial z_u^2}} &=  - \frac{1}{{\sigma _{\rm{p}}^2}}\sum\limits_{i = 1}^M {\int_0^{{T_{\rm{p}}}} {\left( {\frac{{\partial {h_i}}}{{\partial {z_u}}}\frac{{\partial {h_i}}}{{\partial {z_u}}}x_{\rm{p}}^2\left( t \right) + {h_i}\frac{{{\partial ^2}{h_i}}}{{\partial z_u^2}}x_{\rm{p}}^2\left( t \right)} \right.} }\nonumber\\
&\quad\quad\quad\quad\quad\quad\quad\left. { - {y_{{\rm{p}},i}}\left( t \right){x_{\rm{p}}}\left( t \right)\frac{{{\partial ^2}{h_i}}}{{\partial z_u^2}}} \right)\,\mathrm{d}t ,\\
\frac{{{\partial ^2}\Lambda \left( {{{\mathbf{u}}}} \right)}}{{\partial {x_u}\partial {y_u}}}& = \frac{{{\partial ^2}\Lambda \left( {{{\mathbf{u}}}} \right)}}{{\partial {y_u}\partial {x_u}}} = - \frac{1}{{\sigma _{\rm{p}}^2}}\sum\limits_{i = 1}^M {\int_0^{{T_{\rm{p}}}} {\left( {\frac{{\partial {h_i}}}{{\partial {y_u}}}\frac{{\partial {h_i}}}{{\partial {x_u}}}x_{\rm{p}}^2\left( t \right)} \right.} }   \nonumber\\
&\left. { + {h_i}\frac{{{\partial ^2}{h_i}}}{{\partial {x_u}\partial {y_u}}}x_{\rm{p}}^2\left( t \right) - {y_{{\rm{p}},i}}\left( t \right){x_{\rm{p}}}\left( t \right)\frac{{{\partial ^2}{h_i}}}{{\partial {x_u}\partial {y_u}}}} \right)\,\mathrm{d}t, \\
\frac{{{\partial ^2}\Lambda \left( {{{\mathbf{u}}}} \right)}}{{\partial {x_u}\partial {z_u}}} &= \frac{{{\partial ^2}\Lambda \left( {{{\mathbf{u}}}} \right)}}{{\partial {z_u}\partial {x_u}}}= - \frac{1}{{\sigma _{\rm{p}}^2}}\sum\limits_{i = 1}^M {\int_0^{{T_{\rm{p}}}} {\left( {\frac{{\partial {h_i}}}{{\partial {x_u}}}\frac{{\partial {h_i}}}{{\partial {z_u}}}x_{\rm{p}}^2\left( t \right)}  \right.} }  \nonumber\\
&\left. { + {h_i}\frac{{{\partial ^2}{h_i}}}{{\partial {x_u}\partial {z_u}}}x_{\rm{p}}^2\left( t \right) - {y_{{\rm{p}},i}}\left( t \right){x_{\rm{p}}}\left( t \right)\frac{{{\partial ^2}{h_i}}}{{\partial {x_u}\partial {z_u}}}} \right)\,\mathrm{d}t, \\
\frac{{{\partial ^2}\Lambda \left( {{{\mathbf{u}}}} \right)}}{{\partial {y_u}\partial {z_u}}} &= \frac{{{\partial ^2}\Lambda \left( {{{\mathbf{u}}}} \right)}}{{\partial {z_u}\partial {y_u}}} = - \frac{1}{{\sigma _{\rm{p}}^2}}\sum\limits_{i = 1}^M {\int_0^{{T_{\rm{p}}}} {\left( {\frac{{\partial {h_i}}}{{\partial {y_u}}}\frac{{\partial {h_i}}}{{\partial {z_u}}}x_{\rm{p}}^2\left( t \right)}\right.} } \nonumber\\
&\left. { + {h_i}\frac{{{\partial ^2}{h_i}}}{{\partial {y_u}\partial {z_u}}}x_{\rm{p}}^2\left( t \right)  - {y_{{\rm{p}},i}}\left( t \right){x_{\rm{p}}}\left( t \right)\frac{{{\partial ^2}{h_i}}}{{\partial {y_u}\partial {z_u}}}} \right)\,\mathrm{d}t.
\end{align}
\end{subequations}

Since $\mathbb{E}\left\{ {{s_{\text{p}}}} \right\} = 0$, $\mathbb{E}\left\{ {s_{\text{p}}^2} \right\} = \varepsilon$, the expectation of   terms  in  \eqref{partial2} can be  simplified as
\begin{subequations}
\begin{align}
\mathbb{E}\left\{ {\frac{{{\partial ^2}\Lambda \left( {{{\mathbf{u}}}} \right)}}{{\partial x_u^2}}} \right\}
&=  - \frac{{{T_{\rm{p}}}{\left( {P_{\rm{p}}}\varepsilon  + I_{{\rm{DC}}}^2\right)}}}{{\sigma _{\rm{p}}^2}}\sum\limits_{i = 1}^M {\frac{{\partial {h_i}}}{{\partial {x_u}}}\frac{{\partial {h_i}}}{{\partial {x_u}}}} ,\\
\mathbb{E}\left\{ {\frac{{{\partial ^2}\Lambda \left( {{{\mathbf{u}}}} \right)}}{{\partial y_u^2}}} \right\}
&= - \frac{{{T_{\rm{p}}}{\left( {P_{\rm{p}}}\varepsilon  + I_{{\rm{DC}}}^2\right)}}}{{\sigma _{\rm{p}}^2}}\sum\limits_{i = 1}^M {\frac{{\partial {h_i}}}{{\partial {y_u}}}\frac{{\partial {h_i}}}{{\partial {y_u}}}} ,\\
\mathbb{E}\left\{ {\frac{{{\partial ^2}\Lambda \left( {{{\mathbf{u}}}} \right)}}{{\partial z_u^2}}} \right\}
&= - \frac{{{T_{\rm{p}}}{\left( {P_{\rm{p}}}\varepsilon  + I_{{\rm{DC}}}^2 \right)}}}{{\sigma _{\rm{p}}^2}}\sum\limits_{i = 1}^M {\frac{{\partial {h_i}}}{{\partial {z_u}}}\frac{{\partial {h_i}}}{{\partial {z_u}}}} ,\\
\mathbb{E}\left\{ {\frac{{{\partial ^2}\Lambda \left( {{{\mathbf{u}}}} \right)}}{{\partial {x_u}\partial {y_u}}}} \right\}
&= \mathbb{E}\left\{ {\frac{{{\partial ^2}\Lambda \left( {{{\mathbf{u}}}} \right)}}{{\partial {y_u}\partial {x_u}}}} \right\} \nonumber\\&= - \frac{{{T_{\rm{p}}}{\left( {P_{\rm{p}}} \varepsilon + I_{{\rm{DC}}}^2\right)}}}{{\sigma _{\rm{p}}^2}}\sum\limits_{i = 1}^M {\frac{{\partial {h_i}}}{{\partial {x_u}}}\frac{{\partial {h_i}}}{{\partial {y_u}}}} ,\\
\mathbb{E}\left\{ {\frac{{{\partial ^2}\Lambda \left( {{{\mathbf{u}}}} \right)}}{{\partial {x_u}\partial {z_u}}}} \right\}
&= \mathbb{E}\left\{ {\frac{{{\partial ^2}\Lambda \left( {{{\mathbf{u}}}} \right)}}{{\partial {z_u}\partial {x_u}}}} \right\} \nonumber\\&= - \frac{{{T_{\rm{p}}}\left( {P_{\rm{p}}}\varepsilon  + I_{{\rm{DC}}}^2\right)}}{{\sigma _{\rm{p}}^2}}\sum\limits_{i = 1}^M {\frac{{\partial {h_i}}}{{\partial {x_u}}}\frac{{\partial {h_i}}}{{\partial {z_u}}}} ,\\
\mathbb{E}\left\{ {\frac{{{\partial ^2}\Lambda \left( {{{\mathbf{u}}}} \right)}}{{\partial {y_u}\partial {z_u}}}} \right\}
&= \mathbb{E}\left\{ {\frac{{{\partial ^2}\Lambda \left( {{{\mathbf{u}}}} \right)}}{{\partial {z_u}\partial {y_u}}}} \right\} \nonumber\\&= - \frac{{{T_{\rm{p}}}\left({P_{\rm{p}}}\varepsilon  + I_{{\rm{DC}}}^2 \right)}}{{\sigma _{\rm{p}}^2}}\sum\limits_{i = 1}^M {\frac{{\partial {h_i}}}{{\partial {y_u}}}\frac{{\partial {h_i}}}{{\partial {z_u}}}} ,
\end{align}
\end{subequations}
where ${\frac{{\partial {h_i}}}{{\partial {x_u}}}}$,
and ${\frac{{\partial {h_i}}}{{\partial {y_u}}}}$
and ${\frac{{\partial {h_i}}}{{\partial {z_u}}}}$
can be, respectively, expressed as
\begin{subequations}
\begin{align}
  \frac{{\partial {h_i}}}{{\partial {x_u}}}& = \frac{{ - \alpha \left( {m + 3} \right){{\left( {{z_l} - {z_u} - {v_{z,i}}} \right)}^{m + 1}}\left( {{x_u} + {v_{x,i}} - {x_l}} \right)}}{{{{\left\| {{\bf{l}} - {\bf{u}} - {{\bf{v}}_i}} \right\|}^{m + 5}}}},  \\
  \frac{{\partial {h_i}}}{{\partial {y_u}}}& = \frac{{ - \alpha \left( {m + 3} \right){{\left( {{z_l} - {z_u} - {v_{z,i}}} \right)}^{m + 1}}\left( {{y_u} + {v_{y,i}} - {y_l}} \right)}}{{{{\left\| {{\bf{l}} - {\bf{u}} - {{\bf{v}}_i}} \right\|}^{m + 5}}}},\\
  \frac{{\partial {h_i}}}{{\partial {z_u}}}& = \frac{{ - \left( {m + 1} \right)\alpha {{\left( {{z_l} - {z_u} - {v_{z,i}}} \right)}^m}}}{{{{\left\| {{\bf{l}} - {\bf{u}} - {{\bf{v}}_i}} \right\|}^{m + 3}}}} \nonumber\\
  &\quad\quad\quad\quad\quad\quad + \frac{{\left( {m + 3} \right)\alpha {{\left( {{z_l} - {z_u} - {v_{z,i}}} \right)}^{m + 2}}}}{{{{\left\| {{\bf{l}} - {\bf{u}} - {{\bf{v}}_i}} \right\|}^{m + 5}}}}.
\end{align}
\end{subequations}

   \section{Derivation of Equation \eqref{Rate} }\label{tuidao_Rc}

For brevity, we drop the time index $t$ throughout this appendix.
Let $s_1$ and $s_2$ denote values $A$ and $0$, respectively.
According to \eqref{received_data_signal}, the PDF of  ${{y_{{\text{c}}}}}$ can be written as
\begin{align}
f\left( {{y_{\rm{c}}}} \right) = \frac{1}{{2\sqrt {2\pi } {\sigma _{\rm{c}}}}}\sum\limits_{k = 1}^2 {{e^{ - \frac{{{{\left( {{y_{\rm{c}}} - {{\bf{v}}^{\rm{T}}}{\bf{h}}\left( {\sqrt {{P_{\rm{c}}}} {s_k} + {I_{{\rm{DC}}}}} \right)} \right)}^2}}}{{2\sigma _{\rm{c}}^2}}}}}.
\end{align}
Then the mutual information of the receiver is derived as
 \begin{subequations}
\begin{align}
 &I\left( {{x_{\text{c}}};{y_{{\text{c}}}}} \right)
= h\left( {{y_{{\text{c}}}}} \right) - h\left( {{y_{{\text{c}}}}\left| {{x_{\text{c}}}} \right.} \right)\\
 &=  - \int_{ - \infty }^\infty  {f\left( {{y_{\text{c}}}} \right){{\log }_2}f\left( {{y_{\text{c}}}} \right)\,\mathrm{d}{y_{\text{c}}}}  - \frac{1}{2}{\log _2}2\pi e\operatorname{var} \left( {z_{\text{c}}} \right) \\
 & =  - \int_{ - \infty }^\infty  {\frac{{\sum\limits_{k = 1}^2 {{e^{ - \frac{{z_{\rm{c}}^2}}{{2\sigma _{\rm{c}}^2}}}}} }}{{2\sqrt {2\pi } {\sigma _{\rm{c}}}}}{{\log }_2}\frac{{\sum\limits_{j = 1}^2 {{e^{ - \frac{{{{\left( {{{\bf{v}}^{\rm{T}}}{\bf{h}}\sqrt {{P_{\rm{c}}}} \left( {{s_k} - {s_j}} \right) + {z_{\rm{c}}}} \right)}^2}}}{{2\sigma _{\rm{c}}^2}}}}} }}{{2\sqrt {2\pi } {\sigma _{\rm{c}}}}}\,\mathrm{d}} {z_{\rm{c}}} \nonumber\\
 &\quad\quad\quad\quad- \frac{1}{2}{\log _2}2\pi e\sigma _{\rm{c}}^2 \\
 & =  - \frac{1}{2}\int_{ - \infty }^\infty  {\sum\limits_{k = 1}^2 {{f_{{z_{\rm{c}}}}}\left( {{z_{\rm{c}}}} \right)} {{\log }_2}\frac{{\sum\limits_{j = 1}^2 {{e^{ - \frac{{{{\left( {{{\bf{v}}^T}{\bf{h}}\sqrt {{P_{\rm{c}}}} \left( {{s_k} - {s_j}} \right) + {z_{\rm{c}}}} \right)}^2}}}{{2\sigma _{\rm{c}}^2}}}}} }}{{2\sqrt {2\pi } {\sigma _{\rm{c}}}}}\,\mathrm{d}{z_{\rm{c}}}}\nonumber\\
 &\quad\quad\quad\quad- \frac{1}{2}{\log _2}2\pi e\sigma _{\rm{c}}^2\\
 &=   - \frac{1}{2}\sum\limits_{k = 1}^2 \mathbb{E}{{_{{z_{\rm{c}}}}}\left\{ {{{\log }_2}\sum\limits_{j = 1}^2 {\frac{{{e^{ - \frac{{{{\left( {{{\bf{v}}^{\rm{T}}}{\bf{h}}\sqrt {{P_{\rm{c}}}} \left( {{s_k} - {s_j}} \right) + {z_{\rm{c}}}} \right)}^2}}}{{2\sigma _{\rm{c}}^2}}}}}}{2}} } \right\}}\nonumber\\
 &\quad\quad\quad\quad - \frac{1}{{2\ln 2}}.\label{Rc}
\end{align}
\end{subequations}

\section{Derivation of Equation \eqref{Rate_lowerbound}}\label{tuidao_Rc_Ub_Lb}

According to Jensen's Inequality\cite{FiniteAlphabet2012Zeng},
if $f\left( x \right)$ is a convex function, then we have the  inequality
$f\left[ \mathbb{E}{\left( x \right)} \right] \ge \mathbb{E}\left[ {f\left( x \right)} \right]$.
Since ${\log _2}\left( x \right)$ is a concave function with respect to $x$,
according to \eqref{Rc}, a lower bound on the mutual information
is derived as
\begin{subequations}
\begin{align}
&I\left( {{x_{\rm{c}}};{y_{\rm{c}}}} \right)
\ge  - \frac{1}{2}\sum\limits_{k = 1}^2 {{{\log }_2}\sum\limits_{j = 1}^2 {\frac{1}{2}} \mathbb{E}{_{{z_{\rm{c}}}}}\left\{ {{e^{ - \frac{{{{\left( {{{\bf{v}}^{\rm{T}}}{\bf{h}}\sqrt {{P_{\rm{c}}}} \left( {{s_k} - {s_j}} \right) + {z_{\rm{c}}}} \right)}^2}}}{{2\sigma _{\rm{c}}^2}}}}} \right\}} \nonumber\\
&\quad\quad\quad\quad - \frac{1}{{2\ln 2}}\\
& =  - \frac{1}{2}\sum\limits_{k = 1}^2 {{{\log }_2}\sum\limits_{j = 1}^2 {\int_{ - \infty }^\infty  {\frac{{{e^{ - \frac{{{{\left( {{{\bf{v}}^{\rm{T}}}{\bf{h}}\sqrt {{P_{\rm{c}}}} \left( {{s_k} - {s_j}} \right) + {z_{\rm{c}}}} \right)}^2} + z_{\rm{c}}^2}}{{2\sigma _{\rm{c}}^2}}}}}}{2{\sqrt {2\pi } {\sigma _{\rm{c}}}}}\,\mathrm{d}{z_{\rm{c}}}} } }  \nonumber\\
&\quad\quad\quad\quad- \frac{1}{{2\ln 2}}\\
&=  - \frac{1}{2}\sum\limits_{k = 1}^2 {{{\log }_2}\sum\limits_{j = 1}^2 {\frac{1}{{2\sqrt 2 }}} {e^{ - \frac{{{{\left( {{{\bf{v}}^{\rm{T}}}{\bf{h}}\sqrt {{P_{\rm{c}}}} \left( {{s_k} - {s_j}} \right)} \right)}^2}}}{{4\sigma _{\rm{c}}^2}}}}}  - \frac{1}{{2\ln 2}}\\
& =  - {\log _2}\left( {1 + {e^{ - \frac{{{{\left( {{{\bf{v}}^{\rm{T}}}{\bf{h}}} \right)}^2}{P_{\rm{c}}}{A^2}}}{{4\sigma _{\rm{c}}^2}}}}} \right) - \frac{1}{{2\ln 2}} + \frac{3}{2}.
\end{align}
\end{subequations}

Supposed that the bandwidth of the   VLC link is $B$ (Hz). Then, both the input and the output of the VLC link can be represented by samples taken  $\frac{1}{{2B}}$ seconds apart.
Since  the power spectral density of noise is $\frac{{\sigma _{\rm{c}}^2}}{2}$ Watts/Hertz,   the noise power is ${B\sigma _{\rm{c}}^2}$.
For the time interval $\left[ {0,T_{\rm{c}}} \right]$, there are $2BT_{\rm{c}}$ noise samples, and the variance of each sample is $\frac{{B\sigma _{\rm{p}}^2T_{\rm{p}}}}{{2BT_{\rm{c}}}} = \frac{{\sigma _{\rm{c}}^2}}{2}$.
Moreover, if the power of signal is $P$, the signal energy  per sample is $\frac{{PT_{\rm{c}}}}{{2BT_{\rm{c}}}} = \frac{P}{{2B}}$.

Therefore, with  bandwidth  $B$ and ${\bf{h}} = {\bf{\hat h}} + \Delta {\bf{h}}$, the achievable rate is given by
\begin{align}
R_{\rm{c}}^{\rm{L}}\left( {\Delta {\bf{h}}} \right)
 &=3B - \frac{B}{{\ln 2}} - 2B{\log _2}\left( {1 + {e^{ - \frac{{{{\left( {{{\bf{v}}^{\rm{T}}}\left( {{\bf{\hat h}} + \Delta {\bf{h}}} \right)} \right)}^2}{P_{\rm{c}}}{A^2}}}{{4B\sigma _{\rm{c}}^2}}}}} \right).
\end{align}

\end{appendices}

\bibliographystyle{IEEE-unsorted}

\bibliography{refers}

\begin{thebibliography}{10}

\bibitem{Tsonev_2015_Towards}
D.~{Tsonev}, S.~{Videv}, and H.~{Haas},
\newblock ``Towards a 100 {G}b/s visible light wireless access network,''
\newblock {\em Opt. Express}, vol.~23, no.~2, pp.~1627, 2015.

\bibitem{Weichold_2015}
M.~Weichold, M.~Hamdi, M.~Z. Shakir, M.~Abdallah, G.~K. Karagiannidis, and
  M.~Ismail,
\newblock {\em Cognitive Radio Oriented Wireless Networks},
\newblock 2015.

\bibitem{Elgala_2011_Indoor}
H.~Elgala, R.~Mesleh, and H.~Haas,
\newblock ``Indoor optical wireless communication: potential and
  state-of-the-art,''
\newblock {\em IEEE Commun. Mag.}, vol.~49, no.~9, pp.~56--62, Sep. 2011.

\bibitem{Keskin_2018_Localization}
M.~F. Keskin, A.~D. Sezer, and S.~Gezici,
\newblock ``Localization via visible light systems,''
\newblock {\em Proc. IEEE}, vol.~106, no.~6, pp.~1063--1088, Jun. 2018.

\bibitem{Pathak_2015_Visible}
P.~H. Pathak, X.~Feng, P.~Hu, and P.~Mohapatra,
\newblock ``Visible light communication, networking, and sensing: A survey,
  potential and challenges,''
\newblock {\em IEEE Commun. Surv. Tutor.}, vol.~17, no.~4, pp.~2047--2077,
  2015.

\bibitem{Jovicic_2013_Visible}
A.~Jovicic, J.~Li, and T.~Richardson,
\newblock ``Visible light communication: opportunities, challenges and the path
  to market,''
\newblock {\em IEEE Commun. Mag.}, vol.~51, no.~12, pp.~26--32, Dec. 2013.

\bibitem{MA_TWC_2019}
S.~Ma, H.~Li, Y.~He, R.~Yang, S.~Lu, W.~Cao, and S.~Li,
\newblock ``Capacity bounds and interference management for interference
  channel in visible light communication networks,''
\newblock {\em IEEE Trans. Wirel. Commun.}, vol.~18, no.~1, pp.~182--193, Jan.
  2019.

\bibitem{Deng_Access_2019}
X.~Deng, Mardanikorani, G.~Zhou, and J.~Linnartz,
\newblock ``{DC}-bias for optical {OFDM} in visible light communications,''
\newblock {\em IEEE Access}, vol.~7, pp.~98319--98330, 2019.

\bibitem{Gao_TWC_2017}
Qian, Gao, Chen, Gong, Zhengyuan, and Xu,
\newblock ``Joint transceiver and offset design for visible light
  communications with input-dependent shot noise,''
\newblock {\em IEEE Trans. Wirel. Commun.}, vol.~16, no.~5, pp.~2736--2747,
  May. 2017.

\bibitem{2016_Theoretical_Amini}
C.~Amini, A.~Taherpour, T.~Khattab, and S.~Gazor,
\newblock ``Theoretical accuracy analysis of indoor visible light communication
  positioning system based on time-of-arrival,''
\newblock in {\em 2016 IEEE Canadian Conference on Electrical and Computer
  Engineering (CCECE)}, pp. 1--5, 2016.

\bibitem{Akiyama_Time_2017}
T.~Akiyama, M.~Sugimoto, and H.~Hashizume,
\newblock ``Time-of-arrival-based smartphone localization using visible light
  communication,''
\newblock in {\em 2017 International Conference on Indoor Positioning and
  Indoor Navigation(IPIN)}, pp. 1--7, 2017.

\bibitem{Jung_TDOA_2012}
S.~Y. Jung, S.~Hann, and C.~S. Park,
\newblock ``{TDOA}-based optical wireless indoor localization using {LED}
  ceiling lamps,''
\newblock {\em IEEE Trans. Consum. Electron.}, vol.~57, no.~4, pp.~1592--1597,
  Nov. 2011.

\bibitem{2014_Three_Yang}
S.~Yang, H.~Kim, Y.~Son, and S.~Han,
\newblock ``Three-dimensional visible light indoor localization using {AOA} and
  {RSS} with multiple optical receivers,''
\newblock {\em J. Lightwave.Technol.}, vol.~32, no.~14, pp.~2480--2485, Jul.
  2014.

\bibitem{Yu_IPJ_2018}
X.~Yu, J.~Wang, and H.~Lu,
\newblock ``Single {LED}-based indoor positioning system using multiple
  photodetectors,''
\newblock {\em IEEE Photon. J.}, vol.~10, no.~6, pp.~1--8, Dec. 2018.

\bibitem{Bakar_2021_Accurate}
A.~H.~A. Bakar, T.~Glass, H.~Y. Tee, F.~Alam, and M.~Legg,
\newblock ``Accurate visible light positioning using multiple-photodiode
  receiver and machine learning,''
\newblock {\em IEEE Trans. Instrum. Meas.}, vol.~70, pp.~1--12, 2021.

\bibitem{Du_2019_Experimental}
P.~{Du}, S.~{Zhang}, C.~{Chen}, H.~{Yang}, W.~{Zhong}, R.~{Zhang},
  A.~{Alphones}, and Y.~{Yang},
\newblock ``Experimental demonstration of 3{D} visible light positioning using
  received signal strength with low-complexity trilateration assisted by deep
  learning technique,''
\newblock {\em IEEE Access}, vol.~7, pp.~93986--93997, 2019.

\bibitem{Lin_IPJ_2017}
B.~{Lin}, X.~{Tang}, Z.~{Ghassemlooy}, C.~{Lin}, and Y.~{Li},
\newblock ``Experimental demonstration of an indoor {VLC} positioning system
  based on {OFDMA},''
\newblock {\em IEEE Photon. J.}, vol.~9, no.~2, pp.~1--9, Apr. 2017.

\bibitem{Xu_Accuracy_2017}
Y.~{Xu}, Z.~{Wang}, P.~{Liu}, J.~{Chen}, S.~{Han}, C.~{Yu}, and J.~{Yu},
\newblock ``Accuracy analysis and improvement of visible light positioning
  based on {VLC} system using orthogonal frequency division multiple access,''
\newblock {\em Opt. Express}, vol.~25, no.~26, pp.~32618--32630, Dec. 2017.

\bibitem{Yang_Demonstration_2018}
H.~Yang, C.~Chen, W.-D. Zhong, A.~Alphones, S.~Zhang, and P.~Du,
\newblock ``Demonstration of a quasi-gapless integrated visible light
  communication and positioning system,''
\newblock {\em IEEE Photon. Technol. Lett.}, vol.~30, no.~23, pp.~2001--2004,
  Dec. 2018.

\bibitem{Yang_2020_QoS}
H.~Yang, W.-D. Zhong, C.~Chen, A.~Alphones, and P.~Du,
\newblock ``{QoS}-driven optimized design-based integrated visible light
  communication and positioning for indoor iot networks,''
\newblock {\em IEEE Internet Things J.}, vol.~7, no.~1, pp.~269--283, Jan.
  2020.

\bibitem{Yang_WCL_2019}
H.~Yang, P.~Du, W.-D. Zhong, C.~Chen, A.~Alphones, and S.~Zhang,
\newblock ``Reinforcement learning-based intelligent resource allocation for
  integrated {VLCP} systems,''
\newblock {\em IEEE Wireless Commun. Lett.}, vol.~8, no.~4, pp.~1204--1207,
  Aug. 2019.

\bibitem{Yang_Coordinated_2020}
H.~Yang, W.-D. Zhong, C.~Chen, A.~Alphones, P.~Du, S.~Zhang, and X.~Xie,
\newblock ``Coordinated resource allocation-based integrated visible light
  communication and positioning systems for indoor {IoT},''
\newblock {\em IEEE Trans. Wireless Commun.}, vol.~19, no.~7, pp.~4671--4684,
  Jul. 2020.

\bibitem{Keskin_2019_Optimal}
M.~F. Keskin, A.~D. Sezer, and S.~Gezici,
\newblock ``Optimal and robust power allocation for visible light positioning
  systems under illumination constraints,''
\newblock {\em IEEE Trans. Commun.}, vol.~67, no.~1, pp.~527--542, Jan. 2019.

\bibitem{2004Komine}
T.{Komine} and M.{Nakagawa},
\newblock ``Fundamental analysis for visible-light communication system using
  {LED} lights,''
\newblock {\em IEEE Trans. Consum.Electron.}, vol.~50, no.~1, pp.~100--107,
  Feb. 2004.

\bibitem{Wireless1997Kahn}
J.~M. {Kahn} and J.~R. {Barry},
\newblock ``Wireless infrared communications,''
\newblock {\em Proc. IEEE}, vol.~85, no.~2, pp.~265--298, Feb. 1997.

\bibitem{Xu_IPJ_2016}
W.~Xu, J.~Wang, H.~Shen, H.~Zhang, and X.~You,
\newblock ``Indoor positioning for multiphotodiode device using visible-light
  communications,''
\newblock {\em IEEE Photon. J.}, vol.~8, no.~1, pp.~1--11, Feb. 2016.

\bibitem{Simulation_2017_Sindhubala}
K.~Sindhubala and B.~Vijayalakshmi,
\newblock ``Simulation of vlc system under the influence of optical background
  noise using filtering technique,''
\newblock {\em Mater. Today: Proceedings}, vol.~4, no.~2, Part B,
  pp.~4239--4250, 2017.

\bibitem{matlab_fsolve}
 \url{https://ww2.mathworks.cn/help/optim/ug/fsolve.html?s_tid=doc_ta}.

\bibitem{Wang2009}
T.~{Wang}, G.~{Leus}, and L.~{Huang},
\newblock ``Ranging energy optimization for robust sensor positioning based on
  semidefinite programming,''
\newblock {\em IEEE Trans. Signal Process.}, vol.~57, no.~12, pp.~4777--4787,
  Dec. 2009.

\bibitem{Lottici_JSAC_2002}
V.~{Lottici}, A.~{D'Andrea}, and U.~{Mengali},
\newblock ``Channel estimation for ultra-wideband communications,''
\newblock {\em IEEE J. Sel. Areas Commun.}, vol.~20, no.~9, pp.~1638--1645,
  Dec. 2002.

\bibitem{Gezici_SPM_2005}
S.~{Gezici}, Z.~{Tian}, and G.~B. {Giannakis},
\newblock ``Localization via ultrawideband radios,''
\newblock {\em IEEE Signal Process. Mag.}, vol.~22, no.~4, pp.~70--84, Jul.
  2005.

\bibitem{Kay_1993}
S.~M. Kay,
\newblock {\em Fundamentals of Statistical Signal Processing - Estimation
  Theory},
\newblock Englewood Cliffs, NJ: Prentice-Hall, 1993.

\bibitem{Cheung_2004_Least}
K.~W. Cheung, H.~C. So, W.~Ma, and Y.~T. Chan,
\newblock ``Least squares algorithms for time-of-arrival-based mobile
  location,''
\newblock {\em IEEE Trans. Signal Process.}, vol.~52, no.~4, pp.~1121--1130,
  Apr. 2004.

\bibitem{Keskin2018Direct}
M.~F. Keskin, S.~Gezici, and O.~Arikan,
\newblock ``Direct and two-step positioning in visible light systems,''
\newblock {\em IEEE Trans. Commun.}, vol.~66, no.~1, pp.~239--254, Jan. 2018.

\bibitem{Gander1990An}
J.~G. Gander,
\newblock ``An introduction to signal detection and estimation,''
\newblock {\em Signal Process.}, vol.~20, no.~1, pp.~95–96, May. 1990.

\bibitem{Steve2013Distributionally}
S.~Zymler, D.~Kuhn, and B.~Rustem,
\newblock ``Distributionally robust joint chance constraints with second-order
  moment information,''
\newblock {\em Math. Program.}, vol.~137, no.~1, pp.~167--198, Feb. 2013.

\bibitem{Backscatter2019Zhang}
Y.~Zhang, B.~Li, F.~Gao, and Z.~Han,
\newblock ``A robust design for ultra reliable ambient backscatter
  communication systems,''
\newblock {\em IEEE Internet of Things Journal}, vol.~6, no.~5, pp.~8989--8999,
  2019.

\bibitem{Xu2013_SIAM}
Y.~Xu and W.~Yin,
\newblock ``A block coordinate descent method for regularized multiconvex
  optimization with applications to nonnegative tensor factorization and
  completion,''
\newblock {\em SIAM J. Imaging Sci.}, vol.~6, no.~3, pp.~1758--1789, Jan. 2013.

\bibitem{1999Nonlinear}
D.~P. Bertsekas,
\newblock ``Nonlinear programming: 2nd edition,''
\newblock 1999.

\bibitem{Grippof1999}
L.~Grippof and M.~Sciandrone,
\newblock ``Globally convergent block-coordinate techniques for unconstrained
  optimization,''
\newblock {\em Optim. Method Softw.}, vol.~10, no.~4, pp.~587--637, Jan. 1999.

\bibitem{Grippo2000}
L.~Grippo and M.~Sciandrone,
\newblock ``On the convergence of the block nonlinear gauss{\textendash}seidel
  method under convex constraints,''
\newblock {\em Oper. Res. Lett.}, vol.~26, no.~3, pp.~127--136, Apr. 2000.

\bibitem{cvx}
M.~Grant and S.~Boyd,
\newblock ``{CVX}: Matlab software for disciplined convex programming, version
  2.1,'' \url{http://cvxr.com/cvx}, Mar. 2014.

\bibitem{Hu_TSP_2018}
J.~Hu, Y.~Cai, and N.~Yang,
\newblock ``Secure transmission design with feedback compression for the
  internet of things,''
\newblock {\em IEEE Trans. Signal Process.}, vol.~66, no.~6, pp.~1580--1593,
  2018.

\bibitem{Wang_TSP_2015}
H.-M. Wang, C.~Wang, and D.~W.~K. Ng,
\newblock ``Artificial noise assisted secure transmission under training and
  feedback,''
\newblock {\em IEEE Trans. Signal Process.}, vol.~63, no.~23, pp.~6285--6298,
  2015.

\bibitem{Semidefinite2010Luo}
Z.-q. Luo, W.-k. Ma, A.~M.-c. So, Y.~Ye, and S.~Zhang,
\newblock ``Semidefinite relaxation of quadratic optimization problems,''
\newblock {\em IEEE Signal Process. Mag.}, vol.~27, no.~3, pp.~20--34, 2010.

\bibitem{Cartis2010}
C.~Cartis, N.~I.~M. Gould, and P.~L. Toint,
\newblock ``On the complexity of steepest descent, newton{\textquotesingle}s
  and regularized newton{\textquotesingle}s methods for nonconvex unconstrained
  optimization problems,''
\newblock {\em SIAM J. Optim.}, vol.~20, no.~6, pp.~2833--2852, Jan. 2010.

\bibitem{FiniteAlphabet2012Zeng}
W.~{Zeng}, C.~{Xiao}, and J.~{Lu},
\newblock ``A low-complexity design of linear precoding for {MIMO} channels
  with finite-alphabet inputs,''
\newblock {\em IEEE Wireless Commun.Lett.}, vol.~1, no.~1, pp.~38--41, Feb.
  2012.

\end{thebibliography}

\end{document}